\documentclass[fleqn,usenatbib]{mnras}

\usepackage{newtxtext,newtxmath}

\usepackage[T1]{fontenc}

\DeclareRobustCommand{\VAN}[3]{#2}
\let\VANthebibliography\thebibliography
\def\thebibliography{\DeclareRobustCommand{\VAN}[3]{##3}\VANthebibliography}

\usepackage{graphicx}	
\usepackage{amsmath}	
\usepackage{comment}
\usepackage{subcaption}
\usepackage{xcolor}
\usepackage{adjustbox}

\title[Cosmology with GRBs, QSOs, BAO, and SNe Ia]{Gamma-Ray Bursts, Quasars, Baryonic Acoustic Oscillations, and Supernovae Ia: new statistical insights and cosmological constraints}

\author[G. Bargiacchi et al.]{
G. Bargiacchi,$^{1,2}$\thanks{E-mail: giada.bargiacchi@unina.it}
M. G. Dainotti,$^{3,4,5}$\thanks{Corresponding author, maria.dainotti@nao.ac.jp. The first and second author have contributed equally to the paper.}
S. Nagataki, $^{6,7,8}$
S. Capozziello$^{1,2,9}$
\\
$^{1}$Scuola Superiore Meridionale, Largo S. Marcellino 10, 80138 Napoli, Italy \\
$^{2}$Istituto Nazionale di Fisica Nucleare (INFN), Sez. di Napoli, Complesso Univ. Monte S. Angelo, Via Cinthia 9, 80126, Napoli, Italy\\
$^{3}$ National Astronomical Observatory of Japan, 2 Chome-21-1 Osawa, Mitaka, Tokyo 181-8588, Japan\\
$^{4}$ The Graduate University for Advanced Studies, SOKENDAI, Shonankokusaimura, Hayama, Miura District, Kanagawa 240-0193, Japan\\
$^{5}$ Space Science Institute, 4765 Walnut St, Suite B, 80301 Boulder, CO, USA\\
$^{6}$ Interdisciplinary Theoretical \& Mathematical Science Program, RIKEN (iTHEMS), 2-1 Hirosawa, Wako, Saitama, Japan 351-0198\\
$^{7}$ RIKEN Cluster for Pioneering Research, Astrophysical Big Bang Laboratory (ABBL), 2-1 Hirosawa, Wako, Saitama, Japan 351-0198\\
$^{8}$ Astrophysical Big Bang Group (ABBG), Okinawa Institute of Science and Technology Graduate University (OIST),
1919-1 Tancha, Okinawa, Japan 904-0495\\
$^{9}$Dipartimento di Fisica "E. Pancini", Università degli Studi di Napoli Federico II, Complesso Univ. Monte S. Angelo, Via Cinthia 9, 80126, Napoli, Italy
}

\date{Accepted XXX. Received YYY; in original form ZZZ}

\pubyear{2022}

\begin{document}
\label{firstpage}
\pagerange{\pageref{firstpage}--\pageref{lastpage}}
\maketitle

\begin{abstract} 

The recent $\sim 4 \, \sigma$ Hubble constant, $H_{0}$, tension is observed between the value of $H_{0}$ from the Cosmic Microwave Background (CMB) and Type Ia Supernovae (SNe Ia). It is a decade since this tension is excruciating the modern astrophysical community. To shed light on this problem is key to consider probes at intermediate redshifts between SNe Ia and CMB and reduce the uncertainty on $H_0$. 
Toward these goals, we fill the redshift gap by employing Gamma-Ray Bursts (GRBs) and Quasars (QSOs), reaching $z=9.4$ and $z=7.6$, respectively, combined with Baryonic Acoustic Oscillations (BAO) and SNe Ia.
To this end, we employ the ``Dainotti GRB 3D relation"
among the rest-frame end time of the X-ray plateau emission, its corresponding luminosity, and the peak prompt luminosity, 
and the ``Risaliti-Lusso" QSO relation between ultraviolet and X-ray luminosities.
We inquire the commonly adopted Gaussianity assumption on GRBs, QSOs, and BAO.
With the joint sample, we fit the flat $\Lambda$ Cold Dark Matter model with both the Gaussian and the newly discovered likelihoods. We also investigate the impact of the calibration assumed for \textit{Pantheon} and \textit{Pantheon +} SNe Ia on this analysis.  
Remarkably, we show that only GRBs fulfill the Gaussianity assumption.
We achieve
small uncertainties on the matter density parameter $\Omega_M$ and $H_0$. We find $H_0$ values compatible within 2 $\sigma$ with the one from the Tip of the Red Giant Branch. Finally, we show that the cosmological results are heavily biased against the arbitrary calibration choice for SNe Ia.

\end{abstract}


\section{Introduction}

Currently, the spatially flat $\Lambda$ Cold Dark Matter ($\Lambda$CDM) model \citep{peebles1984} is the cosmological model commonly adopted to describe the Universe. It relies on a flat geometry with a cold dark matter (CDM) component and a dark energy corresponding to a cosmological constant $\Lambda$ \citep{2001LRR.....4....1C}. Indeed, this scenario is grounded on several observational probes: the cosmic microwave background \citep[CMB; e.g.][]{planck2018}, the baryonic acoustic oscillations \citep[BAO; e.g.][]{eboss2021}, and the current accelerated expansion of the Universe, discovered from type Ia Supernovae \citep[SNe Ia; e.g.][]{riess1998,perlmutter1999}. Nevertheless, the reliability of this model is still being questioned due to well-known longstanding, but also more recent, theoretical and observational issues: the fine-tuning between the current values of the density parameters of matter ($\Omega_M$) and dark energy ($\Omega_{\Lambda}$), the cosmological constant problem \citep{1989RvMP...61....1W}, the origin and properties of dark energy, and the Hubble constant ($H_0$) tension. The $H_0$ tension is the most recently discovered shortcoming and it corresponds to a discrepancy between the value of $H_0$ measured locally from SNe Ia and Cepheids ($H_0 = 73.04 \pm 1.04  \, \mathrm{km} \, \mathrm{s}^{-1} \, \mathrm{Mpc}^{-1}$, \citealt{2022ApJ...934L...7R}) and the one derived from the Planck data of the CMB with the assumption of a flat $\Lambda$CDM model ($H_0 = 67.4 \pm 0.5  \, \mathrm{km} \, \mathrm{s}^{-1} \, \mathrm{Mpc}^{-1}$, \citealt{planck2018}). This is a $\sim 4 \, \sigma$ deviation, which increases up to 6 $\sigma$ depending on the samples considered \citep{2019ApJ...876...85R,2020PhRvR...2a3028C,2020MNRAS.498.1420W}. Additionally, the investigation of this problem with other cosmological probes has shown a more complex pattern. For example, cosmic chronometers prefer a $H_0$ value close to the one of the CMB \citep{2018JCAP...04..051G}, while time-delay and strong lensing of quasars (QSOs) favor the one of SNe Ia \citep{2019ApJ...886L..23L}, and other probes, such as QSOs \citep{biasfreeQSO2022}, the Tip of the Red-Giant Branch (TRGB; \citealt{2021ApJ...919...16F}), and Gamma-Ray Bursts (GRBs; \citealt{2009MNRAS.400..775C, Postnikov14,2022Galax..10...24D,2022MNRAS.tmp.2639D,2022PASJDainotti}) reveal a $H_0$ value halfway between the two. This challenging scenario has recently boosted the efforts to solve the $H_0$ tension, which still remains one of the most studied and puzzling open questions for the astrophysical and cosmological community (see e.g. \citealt{2022arXiv221116737S}). 

One of the major limitations that prevents to better investigate this discrepancy is the limited number of cosmological probes at redshifts intermediate between the ones of SNe Ia (with a maximum $z=2.26$, \citealt{Rodney}) and CMB ($z=1100$). To fill this gap, thus shedding light on the $H_0$ tension, GRBs and QSOs have been recently applied as high-redshift cosmological tools.

GRBs are extremely powerful and bright sources that are observed up to very high redshifts, reaching $z=8.2$ \citep{2009Natur.461.1254T} and $z=9.4$ \citep{2011ApJ...736....7C}. As a consequence, GRBs could represent a possible new step in the cosmic distance ladder beyond SNe Ia. However, to apply GRBs as probes in cosmology, their physical mechanisms should be first understood. While the community is still debating on the nature of their progenitors and energy processes, their origin is commonly ascribed to two different scenarios. One is the explosion of a massive star \citep{Narayan1992, woosley1993ApJ...405..273W, 1999ApJ...524..262M, 2007ApJ...659..512N,2009ApJ...704..937N,2011PASJ...63.1243N} followed by a core-collapse SNe \citep{2001ApJ...550..410M,2003ApJ...591L..17S} and the other is the merging of two compact objects \citep{1976ApJ...210..549L, 1989Natur.340..126E,1998ApJ...507L..59L}. 
A classification of GRBs based on their observed light curves (LCs) is then crucial to distinguish between the different possible origins. The LCs are generally described with a short prompt high-energy emission followed by an afterglow, which is an emission of longer duration observed in X-ray, optical, and radio wavelengths \citep{1998ApJ...497L..17S,2006ApJ...647.1213O, 2007ApJ...669.1115S,2014ApJ...781...37P,2015ApJ...805...13L,2016ApJ...825L..24M,2017ApJ...835..248W,2018MNRAS.480.4060W}. 
The flat part of GRB LCs after the prompt is the plateau phase, which lasts from $10^2$ to $10^5$ s and is followed by a decay that could be described through a power-law.
An explanation of the plateau can be provided in terms of the external shock model, in which the front of the shock between the interstellar medium and the ejecta is powered by the emission from the central engine \citep{2006ApJ...642..354Z}, or the spin-down of a new-born magnetar \citep{2014MNRAS.443.1779R,2015ApJ...813...92R,2018ApJ...869..155S,2021ApJ...918...12F}.

Historically, GRBs have been classified as Short (SGRBs) and Long (LGRBs), depending on their prompt emission's duration: $T_{90}\leq 2$ s or $T_{90} \ge 2$ s, respectively \citep{1981Ap&SS..80...85M,1993ApJ...413L.101K}, where $T_{90}$ is the time in which a GRB emits from $5\%$ to $95\%$ of the total number of photons emitted in the prompt. This categorization is directly related to the GRB progenitors since LGRBs often derive from the core collapse of a very massive star and SGRBs from the coalescence of two compact objects \citep{2006ApJ...642..354Z,2015ApJ...814L..29I,2017PhRvL.119p1101A,2017Natur.551...71T,2021ApJ...918...59I}. 
Based on the properties of the LCs, several groups have striven to reveal correlations between prompt or plateau, or both GRB features.

Among the correlations involving only plateau properties one can refer to a series of works \citet{Dainotti2008,2010ApJ...725.2209L,2012MNRAS.425.1199B,2012A&A...538A.134X,2013MNRAS.428..729M, Dainotti2013b,2015ApJ...800...31D,2016ApJ...825L..20D, 2016MNRAS.455.1375Z,2019ApJS..245....1T,2019ApJ...883...97Z,2020ApJ...903...18S,2020ApJ...900..168W}.
Some of the previously cited correlations have been applied in cosmology \citep{cardone10, Dainotti2013a,2017NewAR..77...23D,2018PASP..130e1001D,2018AdAst2018E...1D}. 
We here focus on the ``fundamental plane correlation" or the ``3D Dainotti relation" in X-rays \footnote{Our use of the term ``fundamental plane" is strictly related to GRBs and does not refer to other astronomical meanings, such as the fundamental plane of elliptical galaxies \citep{1987ApJ...313...59D}.}. 
This is a three-dimension correlation between the rest-frame time at the end of the plateau emission, $T^{*}_{\chi}$, its X-ray luminosity $L_{\chi}$ \citep{Dainotti2008}, and the prompt peak luminosity, $L_{peak}$ \citep{2016ApJ...825L..20D,2017ApJ...848...88D}. This relation can be theoretically explained within the magnetar model \citep{2011A&A...526A.121D,2012A&A...542A..22B,2014MNRAS.443.1779R} and has already been applied in cosmology \citep{2022MNRAS.514.1828D,2022MNRAS.tmp.2639D} by using a GRB sample properly selected \citep{Dainotti2020a}, as described in Section \ref{data}, and also accounting for the evolution in redshift of the GRB physical quantities (see Section \ref{EP method}).

Concerning QSOs, they are incredibly luminous Active Galactic Nuclei (AGNs) observed up to $z = 7.642$ \citep{2021ApJ...907L...1W}. As for GRBs, their application as high-redshift cosmological probes relies on the understanding of their physical mechanisms, which are still being studied. Indeed, the huge amount of energy emitted from QSOs could not be explained only with standard  stellar processes and galactic components, such as stars, dust, and interstellar gas, but requires an extreme mechanism of different nature. In the commonly accepted picture, QSOs are powered by the accretion on a central supermassive BH, which efficiently converts mass into energy \citep{1998A&A...334...39S,1999RvMPS..71..180H,qsophysics,2020MNRAS.498.5652K}. This scenario can actually reproduce the observed features of QSO emission, especially the ultraviolet (UV) and X-ray emissions. In this regard, a relation between the UV and X-ray luminosities was proposed after the observations of the ﬁrst X-ray surveys
\citep{1979ApJ...234L...9T,1981ApJ...245..357Z,1986ApJ...305...83A}, and then validated by using several different QSO samples \citep[e.g.][]{steffen06,just07,2010A&A...512A..34L,lr16,2021A&A...655A.109B}.
This relation could be ascribed to the above-described QSO engine as follows: the QSO accretion disk around the central supermassive BH produces photons at UV wavelengths, which are processed by an external region of relativistic electrons via the inverse Compton effect, thus originating the emission in X-rays. Besides its viability, this mechanism does not account for the X-ray emission's stability, since the inverse Compton effect should cool down the electrons making them fall onto the central region. To avoid this, a mechanism that efficiently transfers energy from the accretion disk to the external region is required. The origin of such a link is still debated and some theoretical models have been studied to account for the stability of the X-ray emission \citep[see e.g.][]{2017A&A...602A..79L}.

Similarly to GRBs, the application of QSOs in cosmology requires finding correlations among observables intrinsic to the QSO physics.
In principle, the relation between UV and X-ray luminosities just described can be used to infer QSO distances, thus standardizing QSOs as cosmological probes. 

However, the first attempts performed in this direction were strongly limited by the very large intrinsic dispersion of the relation, with a value $ \sim 0.35/0.40$ dex in logarithmic units \citep[][]{2010A&A...512A..34L}. Only recently, it has been discovered that the main contribution to this dispersion is not intrinsic, but due to observational issues \citep{lr16}, thus an accurate selection of the sample is crucial for the cosmological purpose. Carefully removing as many as possible observational biases, such as galaxy contamination, dust reddening, Eddington bias, and X-ray absorption, has led to a reduction of the dispersion up to $\sim 0.2$ dex \citep{2020A&A...642A.150L}. As a consequence, the X-UV relation, denoted as the ``Risaliti-Lusso" (RL) relation in its cosmological application, has turned QSOs into cosmological tools \citep[see e.g.][]{rl15,rl19,2020A&A...642A.150L,2021A&A...649A..65B}.
The reliability of the RL relation has been debated in some recent studies from different points of view. \citet{2022ApJ...935L..19P} have pointed out the presence of the circularity problem in the procedure applied to constrain cosmological parameters with QSOs. In this regard, we here completely overcome this issue following a method already applied to GRBs \citep{2022MNRAS.tmp.2639D} and QSOs \citep{DainottiQSO,biasfreeQSO2022}, as detailed in Section \ref{EP method}. Another point of criticism concerns the possible dependence of the RL relation parameters on both the cosmological model assumed and the redshift \citep{2021MNRAS.502.6140K,2022MNRAS.510.2753K,2022MNRAS.517.1901L}. In this work, we remove the evolution in redshift of luminosities using the same approach already applied to QSOs in \citet{DainottiQSO} and \citet{biasfreeQSO2022} (see Section \ref{EP method}). However, \citet{DainottiQSO} have validated the cosmological use of this relation showing  through well-established statistical tests that it is intrinsic to QSOs and not merely induced by redshift evolution and/or selection effects. Additionally, \citet{biasfreeQSO2022} have also proved that the parameters of the RL relation are compatible within 1 $\sigma$ independently on the cosmological model investigated. 
Thus, the debates on the cosmological applicability of the RL relation are not a concern given the results of \citet{DainottiQSO} and \citet{biasfreeQSO2022} and their innovative approach in the QSO realm, which we apply in this study.

As a matter of fact, the use of GRBs and QSOs could enhance the cosmological analysis both by adding high-$z$ data, more suitable to probe the evolution of the Universe, and by increasing the statistics with additional sources. Indeed, increasing the number of probes leads to smaller uncertainties in the cosmological parameters derived. To this concern, \citet{snelikelihood2022} have recently proved that, as an alternative to the use of huge data sets, it is also possible to significantly reduce the uncertainties on cosmological parameters only through a purely statistical investigation in the probes used. Specifically, they have examined the \textit{Pantheon} \citep{scolnic2018} and \textit{Pantheon +} \citep{pantheon+} SNe Ia samples by inquiring into the statistical assumption generally adopted to constrain cosmological parameters with Gaussian likelihoods. This analysis has shown that both the samples do not verify the Gaussianity assumption and that, by choosing the proper likelihood for each data set, the uncertainties on $\Omega_M$ and $H_0$ are reduced even by a factor $\sim 40 \%$. 
In this epoch dominated by precision cosmology and discrepancies in the measurements of cosmological parameters, constraining them more precisely is key to further understanding, and even alleviating or solving, these tensions. 

To this end, we here leverage the advantages of both the possible approaches described above: the inclusion of high-redshift probes in the cosmological analysis and the examination of the statistical foundation of each probe considered. Indeed, in this study, we investigate the Gaussianity assumption and uncover the proper cosmological likelihood for the samples of BAO, GRBs, and QSOs. Then, we combine these data with the SNe Ia samples from the \textit{Pantheon} and \textit{Pantheon +} releases to fit a flat $\Lambda$CDM model by using, in one case, the standard method with Gaussian likelihoods and, in the other case, the novel discovered likelihoods. In both cases, GRB and QSO redshift evolution are treated following different approaches. In addition, we delve into the a-priori assumptions imposed by the \textit{Pantheon} and \textit{Pantheon +} releases to test to what extent these assumptions affect the cosmological results. This work presents clear points of originality compared to other previous analyses in this realm. 
Although GRBs, QSOs, SNe Ia and BAO have already been applied jointly \citep[e.g.][]{2021arXiv211107907S}, this has never been done with the GRB fundamental plane relation and with the most updated samples of SNe Ia, GRB, and QSO samples.
This is also the first time that the Gaussianity assumption is inspected and the actual proper cosmological likelihoods are uncovered for the investigated samples of BAO, GRBs, and QSOs. This analysis allows us to investigate if and to what extent the use of different likelihoods impacts the best-fit values and the uncertainties of $\Omega_M$ and $H_0$ in a flat $\Lambda$CDM model, also in light of the comparison of our results with determinations of these parameters from other sources. 
We also stress here that the main goal of our study is to prove that QSOs and GRBs can be implemented in the cosmological analysis to extend the redshift range up 7.5 without introducing additional uncertainties on the cosmological parameters also with the new likelihood functions used. This is the first essential step to enhance their use in cosmology. The improvement of the cosmological power of QSOs and GRBs is underway, but it is out of the scope of this paper.

The manuscript is structured as follows. Section \ref{data} describes each sample used and its selection. Section \ref{methodology} details the methods employed for all the steps of our analysis. The treatment of the redshift evolution, selection biases, and circularity problem for GRBs and QSOs is in Section \ref{EP method}, the physical quantities investigated for each probe is in Section \ref{quantities}, the normality tests is applied in Section \ref{normalitytests}, and the search for the best-fit likelihoods, the fit of the flat $\Lambda$CDM model, and the test on SNe Ia assumptions are in Section \ref{newfits}. Section \ref{results} reports the results of all our investigations. The outcomes of the tests on Gaussianity and the new likelihoods are presented in Section \ref{normalresults} and Section \ref{bestfitlikelihood}, respectively, while the cosmological findings and implications are detailed in Section \ref{cosmologicalfits} together with the comparison with previous works in the literature. We summarize our work and draw our conclusions in Section \ref{conclusions}.

\section{Data}
\label{data}

This study makes use of the combination of four cosmological probes: SNe Ia, GRBs, QSOs, and BAO. We here detail the data set considered for each of them.

We employ the two most recent SNe Ia collections of ``\textit{Pantheon}" \citep{scolnic2018} and ``\textit{Pantheon +}" \citep{pantheon+} with the full covariance matrix that includes both statistical and systematic uncertainties. The first sample is composed of 1048 sources in the redshift range between $z=0.01$ and $z=2.26$ gathered from CfA1-4, \textit{Carnegie Supernova Project}, \textit{Pan-STARRS1},  \textit{Sloan Digital Sky Survey} (SDSS), \textit{Supernova Legacy Survey}, and \textit{Hubble Space Telescope}, while the second one consists of 1701 SNe Ia collected from 18 different surveys in the range $z=0.001 - 2.26$. The \textit{Pantheon +} sample significantly enhances the previous release with an increased redshift span and number of sources, in particular at low redshift, and with an enriched treatment of systematic uncertainties. These improvements enable SNe Ia to better constrain cosmological parameters \citep[see e.g.][]{2022ApJ...938..110B}. The use of both samples in this work allows us to reveal if and how our analysis is affected by the choice of the SNe Ia sample.

For GRBs, we use the so-called ``Platinum" sample \citep{2020ApJ...904...97D} composed of 50 X-ray sources between $z=0.055$ and $z=5$ that have been selected according to the following procedure \citep[see also][]{2022MNRAS.tmp.2639D,2022MNRAS.514.1828D}. Starting from the 372 GRBs observed by the {\it Neil Gehrels Swift Observatory} (Swift) from January 2005 to August 2019 with a known redshift from Swift+Burst Alert Telescope (BAT)+ X-Ray Telescope (XRT) repository \citep{Evans2009}, the 222 which present a reliable plateau and which can be fitted using the \cite{2007ApJ...662.1093W} model are retained. To reduce the intrinsic dispersion of the 3D Dainotti relation and restrict to a more homogeneous physical mechanism, only LGRBs are considered
\citep[see][for details]{2016ApJ...825L..20D,2017ApJ...848...88D, 2020ApJ...904...97D}. The final sub-sample is then obtained by removing sources that present 1) a plateau that lasts $<500$ s, or 2) less than 5 points in the region before the plateau, or 3) a time at the end of the plateau that could not be directly determined as it falls within an observational gap region, or 4) flares in the plateau, or 5) a not well-defined starting point of the plateau phase, or 6) a plateau inclination $>$41° \citep[see also][]{2016ApJ...825L..20D,2017ApJ...848...88D,2020ApJ...904...97D}. These criteria define our final GRB sample.

The QSO data set is the one detailed in \citet{2020A&A...642A.150L}, with 2421 sources from $z=0.009$ to $z =7.54$ \citep{banados2018} and assembled by eight samples from
literature and archives. Specifically, these data belong to \citet{2019A&A...632A.109N}, \citet{salvestrini2019}, and \citet{2019A&A...630A.118V} samples, XMM–XXL North QSO sample \citep{2016yCat..74570110M}, SDSS Data Release 14 (SDSS DR14; \citealt{2018A&A...613A..51P}), 4XMM Newton \citep{2020A&A...641A.136W}, and \textit{Chandra} \citep{2010ApJS..189...37E}, with additional low-$z$ QSOs with UV measurements from the International Ultraviolet Explorer and X-ray archival data. Lots of efforts have been spent to select these sources to be suitable for cosmological studies by carefully removing as much as possible observational biases, as described in \citet{rl15}, \citet{lr16}, \citet{rl19}, \citet{salvestrini2019} and \citet{2020A&A...642A.150L}. We below summarize the main criteria applied in this selection procedure.
As a preliminary screening, sources with a signal-to-noise ratio S/N $< 1$ are discarded. Then, only QSOs with extinction E(B-V)$\leq 0.1$ are retained to remove UV reddening and contamination of the host galaxy. The corresponding requirement to be satisfied is $\sqrt{(\Gamma_{1,\mathrm{UV}}-0.82)^2 + (\Gamma_{2,\mathrm{UV}}-0.40)^2} \leq 1.1$, where $\Gamma_{1, \mathrm{UV}}$ and $\Gamma_{2, \mathrm{UV}}$ are the slopes of the log($\nu$)-log($\nu \, L_\nu$) power-law in the rest-frame 0.3-1$\mu$m and 1450-3000 \AA \, ranges, respectively, and $L_\nu$ is the luminosity per unit of  $\nu$. The values  $\Gamma_{1, UV}=0.82$ and $\Gamma_{2, UV}=0.4$ correspond to a spectral energy distribution (SED) with zero extinction \citep[see][]{2006AJ....131.2766R}. The possible presence of absorption in X-ray is then accounted for by requiring $\Gamma_X + \Delta \Gamma_X \geq 1.7$ and $\Gamma_X \leq 2.8 $ if $z < 4$ and $\Gamma_X \geq 1.7$ if $z \geq 4$, where $\Gamma_X$ is the photon index and $\Delta \Gamma_X$ its associated uncertainty. The last step of this selection procedure consists of correcting QSOs for the Eddington bias by imposing $\text{log}F_{X,exp} - \text{log}F_{min} \geq {\cal F}$, where $F_{X,exp}$ is the X-ray flux computed from the UV flux with the assumptions of the RL relation and a flat $\Lambda${CDM} model with $\Omega_{M}=0.3$ and $H_{0} = 70\, \mathrm{km\,s^{-1}\,Mpc^{-1}}$. $F_{min}$ represents the minimum detectable flux of the observation that, as detailed in \citet{lr16}, can be computed for each source from the total exposure time of the charge-coupled device (CCD), provided in the catalogues, by using the functions presented by \citet{2001A&A...365L..51W} in Fig. 3. The threshold value ${\cal F}$ is ${\cal F} = 0.9$ for the SDSS DR14– 4XMM Newton and XXL sub-samples and ${\cal F} = 0.5$ for the SDSS DR14-\textit{Chandra}. Since multiple X-ray observations of the same QSO can survive these filters, they are averaged, thus reducing also the X-ray variability effects.
As in \citet{DainottiQSO} and \citet{biasfreeQSO2022}, we use this final cleaned sample without imposing any cut in redshift, such as the one at $z=0.7$ investigated in \citet{2020A&A...642A.150L}, to avoid introducing arbitrary truncation.

The BAO collection we use is the one described in \citet{2018arXiv180707323S}, composed of 26 data points for which the covariance matrix is detailed in \citet{2016JCAP...06..023S}. Out of these 26, 24 measurements are in the low redshift range between $z=0.106$ \citep{2011MNRAS.416.3017B} and $z=0.73$ \citep{2011MNRAS.418.1707B}, with data from \citet{2012MNRAS.427.2132P}, \citet{2013MNRAS.435..255C}, \citet{2013MNRAS.433.3559C}, \citet{2014MNRAS.441...24A}, \citet{2015MNRAS.449..835R}, and \citet{2017MNRAS.469.3762W}, while the other two measurements are at $z=2.34$ \citep{2015A&A...574A..59D} and $z=2.36$ \citep{2014JCAP...05..027F}. This data set has already been used in \citet{2022Galax..10...24D} and \citet{2022MNRAS.tmp.2639D} in combination with \textit{Pantheon} SNe Ia and  GRBs.

\section{Methodology}
\label{methodology}

\subsection{Redshift evolution treatment}
\label{EP method}

GRBs and QSOs are high-redshift sources, which establishes their invaluable role in investigating the evolution of the Universe. Nevertheless, at high redshifts we need to account for possible selection biases and evolutionary effects, that can in principle distort or even induce a correlation between physical quantities of a source, leading to an incorrect determination of the cosmological parameters \citep{Dainotti2013a}. To correct for these effects, we can use the Efron \& Petrosian (EP) statistical method \citep{1992ApJ...399..345E}, which has already been applied to the GRB \citep{Dainotti2013a,Dainotti2015b,2017A&A...600A..98D,2021Galax...9...95D,2022MNRAS.tmp.2639D} and QSO domains \citep{DainottiQSO,biasfreeQSO2022}. In this work, we make use of the results reported in \citet{2022MNRAS.tmp.2639D} for GRBs and in \citet{DainottiQSO} and \citet{biasfreeQSO2022} for QSOs. We here briefly summarize their methods and outcomes.

In their approach, the physical quantities of interest, luminosities and also time in the case of GRBs, evolve with redshift as $L' = \frac{L}{(1+z)^{k}}$ and $T' = \frac{T}{(1+z)^{k}}$, where $L$ and $T$ are the observed quantities, $L'$ and $T'$ the corresponding corrected ones, and $k$ the evolutionary parameter. As proved in \citet{2011ApJ...743..104S}, \citet{2021ApJ...914L..40D}, and \citet{DainottiQSO}, the results of this method are not affected by the choice of the power-law, which could also be replaced by more complex functions of the redshift. 
To determine the value of $k$ that removes the dependence on the redshift, the Kendall's $\tau$ statistic is applied, where the coefficient $\tau$ is defined as
\begin{equation}
\label{tau}
    \tau =\frac{\sum_{i}{(\mathcal{R}_i-\mathcal{E}_i)}}{\sqrt{\sum_i{\mathcal{V}_i}}}.
\end{equation}
Here, the index $i$ refers to all the sources that at redshift $z_i$ have a luminosity greater than the lowest luminosity ($L_{min,i}$) that can be observed at that redshift. This minimum luminosity is computed by choosing a limiting flux. The assumed value must guarantee that the retained sample is at least 90\% of the total one and that it resembles as much as possible the overall
distribution, which can be verified by applying the Kolmogorov-Smirnov test \citep{Dainotti2013a,Dainotti2015b,2017A&A...600A..98D,2022ApJ...925...15L,DainottiQSO,2022MNRAS.514.1828D}. The rank $\mathcal{R}_i$ in Eq. \eqref{tau} corresponds to the number of data points in the associated set of the $i$-source, where the associated set is defined by all $j$-points that verify $z_j \leq z_i$ and $L_{z_j} \geq  L_{min,i}$. $\mathcal{E}_i = \frac{1}{2}(i+1)$ and $\mathcal{V}_i = \frac{1}{12}(i^{2}+1)$ are the expectation value and variance, respectively, in the case of the absence of correlation. Thus, the redshift dependence is removed when $\tau = 0$, and this condition provides us the value of $k$ that eliminates the correlation. If $| \tau | > n$ the hypothesis of uncorrelation is rejected at $n \sigma$ level, hence we obtain the 1 $\sigma$ uncertainty on the $k$ value by requiring $|\tau| \leq 1$. The uncovered dependence on the redshift is then used to derive the de-evolved $L'$ for the total sample. This procedure can be straightforward applied also to the time variable for GRBs by replacing the luminosity with the time.
The $k$ values obtained for the quantities of our interest and used in this work are $k_{L_{peak}} = 2.24 \pm 0.30$, $k_{T_{\chi}} = - 1.25^{+0.28}_{-0.27}$, and $k_{L_{\chi}} = 2.42^{+0.41}_{-0.74}$ for GRBs \citep{2022MNRAS.tmp.2639D}, and $k_{UV} = 4.36 \pm 0.08$ and $k_X = 3.36 \pm 0.07$ for QSOs \citep{DainottiQSO}. The de-evolved quantities computed with these $k$ values are thus used in our cosmological fits when accounting for a ``fixed" evolution. The notation ``fixed" refers to the fact that $k$ is determined under the assumption of a specific cosmological model, which is required to compute the luminosities $L$ from the measured fluxes. In both \citet{2022MNRAS.tmp.2639D} and \citet{DainottiQSO} the fiducial model is a flat $\Lambda$CDM model with $\Omega_M =0.3$ and $H_0 = 70  \, \mathrm{km} \, \mathrm{s}^{-1} \, \mathrm{Mpc}^{-1}$. As a consequence, this method suffers from the ``circularity problem": the a-priori assumption of a cosmological model affects the constraints on cosmological parameters that are found by fitting the luminosities computed under this assumption.
To overcome this issue, \citet{2022MNRAS.tmp.2639D} (for GRBs), and \citet{DainottiQSO} and \citet{biasfreeQSO2022} (for QSOs) have studied the behaviour of $k$ as a function of the cosmology. Specifically, in these works, $k$ is evaluated not for fixed cosmological parameters, but over a set of several values of the cosmological parameters (i.e. $\Omega_M$, $H_0$, and also other parameters for extensions of the flat $\Lambda$CDM model), leading to the functions $k(\Omega_M)$ and $k(H_0)$. These analyses show that $k$ does not depend on $H_0$, while it depends on $\Omega_M$. Hence, the function $k(\Omega_M)$ can be applied in the cosmological fits to let $k$ vary together with the cosmological parameters, without the need to fix a cosmology. This method completely solves the circularity problem and we refer to it as ``varying" evolution. In this work, we compare the results from all possible treatments of the evolution: without correction, with ``fixed", and ``varying" evolution (see Table \ref{tab:bestfit}).
We are aware that the a more complete and independent treatment of the evolution would require to leave free to vary the sets of four parameters for the evolution, $k$ for $L_{X, QSOs}$, $L_{UV, QSOs}$, $L_{X,GRBs}$, $L_{peak,GRBs}$. However, we here anticipate that if we apply this more general procedure imposing uniform priors on the evolutionary parameters, we cannot constrain the $k$ parameters. Thus, the only chance to obtain an idea of how the evolution would play a role is to impose Gaussian priors on the evolution based on the analysis of how $k(\Omega_M)$ varies for the luminosities. We perform these fits by imposing a Gaussian prior in which the mean corresponds to the expected value of $k$ and the standard deviation is five times the error on this $k$ value. More precisely, the values of $k$ assumed as mean of the Gaussian and their errors are $k_{L_{peak,GRBs}} = 2.24 \pm 0.30$ and $k_{L_{X,GRBs}} = 2.42 \pm 0.58$ for GRBs, and $k_{L_{UV,QSOs}} = 4.36 \pm 0.08$ and $k_{L_{X,QSOs}} = 3.36 \pm 0.07$ for QSOs, when considering the \textit{Pantheon} sample, while $k_{L_{peak,GRBs}} = 2.19 \pm 0.29$ and $k_{L_{X,GRBs}} = 2.37 \pm 0.56$ for GRBs, and $k_{L_{UV,QSOs}} = 4.29 \pm 0.08$ and $k_{L_{X,QSOs}} = 3.29 \pm 0.06$ for QSOs, when using the \textit{Pantheon +} SNe Ia. These values are indeed the ones obtained from the functions $k(\Omega_M$) when assuming $\Omega_M=0.3$ for \textit{Pantheon} and $\Omega_M=0.35$ for \textit{Pantheon +}. The results of these fits are reported in Figures \ref{fig: kfree Gaussian} and \ref{fig: kfree newlikelihoods} for all cases studied in this work.
\subsection{Theoretical and measured physical quantities}
\label{quantities}

SNe Ia presents an almost uniform intrinsic luminosity and thus can be considered among the most reliable classes of \textit{standard candles}. To apply them as cosmological probes, we use the distance modulus $\mu$ defined as $\mu = m - M$, where $m$ and $M$ are the apparent and absolute magnitude, respectively. The critical point in this definition is the computation of $M$ that depends on several factors, such as selection biases, microlensing, and different contributions to statistical and systematic uncertainties \citep{scolnic2018}. The observed SNe Ia $\mu$ is 
\begin{equation}
\label{muobs}
\mu_{obs, \mathrm{SNe \, Ia}} = m_B - M + \alpha x_1 - \beta c + \Delta_M + \Delta_B
\end{equation}
where $m_B$ is the B-band overall amplitude, $x_1$ the stretch parameter, $c$ the color, $\alpha$ and $\beta$ the coefficients of the relation of luminosity with $x_1$ and $c$, respectively, $M$ the fiducial B-band absolute magnitude of a SN with $x_1 = 0$ and $c=0$, $\Delta_M$ and $\Delta_B$ the corrections on the distance that account for the mass of the host-galaxy and biases predicted through simulations, respectively \citep{scolnic2018,pantheon+}.  $M$ is degenerate with $H_0$, hence $H_0$ cannot be determined by SNe Ia alone
\citep{1998A&A...331..815T,scolnic2018}.
However, $H_0$ can be derived if $M$ is fixed.
In this regard, we employ the data directly provided by \textit{Pantheon} and \textit{Pantheon +}, thus we do not fix a priori ourselves any value of $M$, but we use the same assumption of \textit{Pantheon} and values of \textit{Pantheon +} releases.
Indeed, primary distance anchors, such as Cepheids and the TRGB, are usually used to calibrate $M$ from which $\mu$ is then computed \citep{2022ApJ...938..110B,2022ApJ...934L...7R}. \citet{scolnic2018} fix $M=-19.35$
corresponding to $H_0 = 70 \, \mathrm{km} \, \mathrm{s}^{-1} \, \mathrm{Mpc}^{-1}$ \citep[][]{2021ApJ...912..150D,2022Galax..10...24D}. \citet{2022ApJ...934L...7R} find $M = -19.253 \pm 0.027$ and $H_0 = 73.04 \pm 1.04  \, \mathrm{km} \, \mathrm{s}^{-1} \, \mathrm{Mpc}^{-1}$ using 42 SNe Ia combined with Cepheids hosted in the same galaxies of these SNe Ia. The $\mu$ provided for the \textit{Pantheon +} sample are computed assuming this value of $M$. 
In our analysis, we directly use the $\mu_{obs, \mathrm{SNe \, Ia}}$ supplied by \textit{Pantheon}\footnote{\url{https://github.com/dscolnic/Pantheon}} and \textit{Pantheon +}\footnote{\url{https://github.com/PantheonPlusSH0ES}} releases. 
Within a flat $\Lambda$CDM model and ignoring the current relativistic density parameter, whose contribution is negligible in the late epochs, the theoretical $\mu$ is
\begin{equation}
\label{dmlcdm}
\mu_{th} = 5 \, \mathrm{log_{10}} \, d_l (\Omega_M, H_0) + 25 = 5 \, (1+z) \frac{c}{H_{0}} \, \int^{z}_{0} \frac{dz'}{E(z')} + 25
\end{equation}
where $d_l$ is the luminosity distance in Megaparsec (Mpc), $c$ the speed of light, and $E(z)$ the dimensionless Hubble parameter given by
\begin{equation}
\label{E(z)}
E(z) = \frac{H(z)}{H_{0}} = \sqrt{\Omega_{M} (1+z)^{3} + (1- \Omega_{M})}.
\end{equation}
In the case of SNe Ia, we slightly modify Eq. \eqref{dmlcdm} using the more precise formula provided by \citet{2019ApJ...875..145K}
\begin{equation}
\label{dmlcdm_corr}
\mu_{th} = 5 \, (1+z_{hel}) \frac{c}{H_{0}} \, \int^{z_{HD}}_{0} \frac{d z'}{\sqrt{\Omega_{M} (1+z)^{3} + (1- \Omega_{M})}} + 25
\end{equation}
where $z_{HD}$ the ``Hubble-Lema\^itre diagram” redshift that accounts for peculiar velocity and CMB corrections and $z_{hel}$ is the heliocentric redshift.\\

For GRBs, the 3D X-ray fundamental plane relation employed to use them as cosmological tools has the following form:
\begin{equation}
\log_{10} L_{\chi} =  a \, \log_{10} T^{*}_{\chi} + b \, \log_{10} L_{peak} + c.
\label{3drelation}
\end{equation}
The luminosities are computed from the measured fluxes by applying the relation between fluxes $F$ and luminosities $L$, $L= 4 \pi d_{l}^{2}\, \cdot  F \cdot  \, K$, where $d_l$ is in units of cm and $K$ is K-correction accounting for the cosmic expansion \citep{2001AJ....121.2879B}. As the GRB spectrum is well reproduced by a simple power-law, the K-correction is given by $ K = (1+z)^{\gamma -1}$, with $\gamma$ the spectral index of the X-ray plateau.
The values of $a$, $b$, and $c$ parameters in Eq. \eqref{3drelation} are determined by fitting the 3D relation with the Bayesian technique of the D'Agostini method \citep{2005physics..11182D}, which is based on the Markov Chain Monte Carlo (MCMC) approach and allows us to consider error bars on all quantities and also an intrinsic dispersion $sv$ of the relation. 
As detailed in Section \ref{EP method}, $L_{\chi}$,  $T^{*}_{\chi}$, and $L_{peak}$ in Eq. \eqref{3drelation} could also be corrected to account for their redshift evolution through the application of the EP method \citep{2022MNRAS.tmp.2639D}.
As for SNe Ia, the physical quantity we consider for GRBs is $\mu$. The observed $\mu$ for GRBs ($\mu_{obs, \mathrm{GRBs}}$) is computed assuming the 3D relation. Converting luminosities into fluxes in Eq. \eqref{3drelation} through the above-defined relation between these two quantities and using $\mu = 5 \, \mathrm{log_{10}} \, d_l + 25$ \citep[see][for the mathematical derivation]{2022MNRAS.tmp.2639D}, we obtain 
\begin{small}
\begin{equation}
\begin{split}
\label{dmGRBs}
\mu_{obs, \mathrm{GRBs}} = & 5 \Bigg[ -\frac{\log_{10} F_{\chi}}{2 (1-b)}+\frac{b \cdot \log_{10} F_{peak}}{2 (1-b)} - \frac{(1-b)\log_{10}(4\pi)+c}{2 (1-b)}+ \\ 
& + \frac{a \log_{10} T^{*}_{\chi}}{2 (1-b)} \Bigg] + 25
\end{split}
\end{equation}
\end{small}
where we consider the K-correction already applied to all quantities. Following \citet{2022MNRAS.tmp.2639D}, we fix $c=23$ when applying Eq. \eqref{dmGRBs}.
Under the same cosmological assumption applied to SNe Ia, $\mu_{th}$ for GRBs is the one already defined in Eq. \eqref{dmlcdm} with $E(z)$ specified in Eq. \eqref{E(z)} .\\

To standardize QSOs as cosmological probes and compute their distances, we use the X-UV RL relation, which is commonly parameterized through a linear relation between logarithmic quantities as
\begin{equation}\label{RL}
\mathrm{log_{10}} \, L_{X} = g_1 \, \mathrm{log_{10}} \, L_{UV} + n_1
\end{equation}
where $L_X$ and $L_{UV}$ are the luminosities (in units of $\mathrm{erg \, s^{-1} \, Hz^{-1}}$) at 2 keV and 2500 \AA, respectively. 
As for GRBs, we compute $L_{X}$ and $L_{UV}$ in Eq. \eqref{RL} from the observed flux densities $F_{X}$ and $F_{UV}$ (in $\mathrm{erg \, s^{-1} \, cm^{-2} \, Hz^{-1}}$), respectively, using $L_{X,UV}= 4 \pi d_{l}^{2}\ F_{X, UV}$. Indeed, for QSOs the spectral index $\gamma$ is assumed to be 1, leading to $K=1$ \citep{2020A&A...642A.150L}, and thus the K-correction is omitted in this case. As detailed in Section \ref{EP method} and already stressed for GRBs, we can correct $L_X$ and $L_{UV}$ for evolutionary effects by applying the EP method \citep{DainottiQSO,biasfreeQSO2022}. The parameters $g_1$ and $n_1$ can be fitted, along with the intrinsic dispersion $sv_1$ of the RL relation, through the D'Agostini method.
Inserting $\mathrm{log_{10}}L_{UV} =  \mathrm{log_{10}}(4 \, \pi \, d_l^2) + \mathrm{log_{10}}F_{UV}$ in Eq. \eqref{RL} provides us the observed physical quantity $\mathrm{log_{10}}L_{X,obs}$ under the assumption of the RL relation:
\begin{equation}
\label{lxobs}
\mathrm{log_{10}} \, L_{X,obs} = g_1 \left[ \mathrm{log_{10}}(4 \, \pi \, d_l^2) + \mathrm{log_{10}}F_{UV} \right] + n_1.
\end{equation}
The theoretical quantity is computed according to $\mathrm{log_{10}}L_{X,th} = \mathrm{log_{10}}(4 \, \pi \, d_l^2) + \mathrm{log_{10}}F_X$. Both $\mathrm{log_{10}}L_{X,obs}$ and $\mathrm{log_{10}}L_{X,th}$ require a fixed cosmological model in the computation of $d_l$.

To be more precise, in principle, we could also consider as physical quantities the logarithmic luminosities for GRBs and the distance moduli for QSOs, since the approaches with luminosities and with distance moduli are equivalent (i.e. they are related through $d_l$). Nevertheless, we have here described only the case of distance moduli for GRBs and the case of luminosities for QSOs to be consistent with the fitting procedure we apply in the following cosmological analysis. Concerning GRBs, \citet{2022MNRAS.tmp.2639D} have proved that the cosmological constraints do not depend on the choice of using distance moduli or logarithmic luminosities in the likelihood; indeed, the values of cosmological parameters obtained in the two approaches are consistent within 1 $\sigma$. Thus, we prefer to use the distance moduli to remove one degree of freedom, as the parameter $c$ is fixed in this case \citep{2022MNRAS.tmp.2639D}
as this choice also guarantees the same number of free parameters (2) for both the 3D and the RL relations. Concerning QSOs, we prefer to construct the cosmological likelihood with the logarithmic luminosities as these are the quantities intrinsic to the RL relation (see Eq.\eqref{RL}) and also the ones that are commonly and robustly used for cosmological analyses in the literature \citep[e.g.][]{2020MNRAS.492.4456K,2020MNRAS.497..263K,2020A&A...642A.150L,2021MNRAS.502.6140K,2022arXiv220611447C,2022MNRAS.510.2753K,2022arXiv220310558C,2022MNRAS.515.1795B,biasfreeQSO2022,2022arXiv221014432W,2022MNRAS.517.1901L}. Thus, this approach allows an easier and more immediate comparison with results from other studies, without the need for taking into account possible differences in the respective methodologies. 

For BAO, the investigated physical quantity is $d_{z} = r_s(z_d)/D_V(z)$ \citep[e.g.][]{2005ApJ...633..560E}, where $r_s(z_d)$ is the sound horizon at the baryon drag epoch $z_d$ and $D_V(z)$ is the volume averaged distance. The observed $d_{z,obs}$ provided in \citet{2018arXiv180707323S} are obtained from the measured $D_V(z)$ and assuming the fiducial $(r_s(z_d) \, \cdot h)_{\mathrm{fid}} = 104.57$ Mpc, where $h$ is the dimensionless Hubble constant $\displaystyle h= {H_{0}}/{100 \, \mathrm{km \,s^{-1} \, Mpc^{-1}}}$, which corresponds to the best-fit of a $\Lambda$CDM model \citep[see][]{2016JCAP...06..023S}.
The theoretical $d_{z,th}$ are instead computed as follows. Since an exact computation of $r_s(z_d)$ would require the use of Boltzmann codes, we estimate it with the following numerical approximation \citep{2015PhRvD..92l3516A,2019JCAP...10..044C}:

\begin{equation}
\label{rs}
r_s(z_d) \sim \frac{55.154 \, e^{-72.3  (\Omega_{\nu}\, h^2 + 0.0006)^{2}}}{(\Omega_M \, h^2)^{0.25351} \, (\Omega_b \, h^2)^{0.12807}} \, \mathrm{Mpc}
\end{equation}
where $\Omega_{b}$ and $\Omega_{\nu}$ are the baryon and neutrino density parameters. In this formula, we fix $\Omega_{\nu} \, h^2 = 0.00064$ and $\Omega_{b} \, h^2 = 0.002237 $ (according to \citealt{Hinshaw_2013} and \citealt{planck2018}).
The theoretical distance $D_V(z)$ required to compute $d_{z,th}$ is defined as \cite{2005ApJ...633..560E}
\begin{equation} \label{DVBAO}
D_{V}(z) = \left[ \frac{cz}{H(z)} \frac{d_l^{2}(z)}{(1+z)^{2}} \right]^{\frac{1}{3}}.
\end{equation}

\subsection{Tests of the Gaussianity assumption}
\label{normalitytests}

To check the assumptions underlying the applicability of a Gaussian likelihood ($\cal L$$_{Gaussian}$) to fit cosmological models, we apply to each probe several methods to test if the difference between the measured and the theoretical quantities is normally distributed. Indeed, this is the statistical condition required to constrain cosmological parameters by applying the likelihood function 
\begin{equation}
\label{Lgauss}
\mathcal{L}_{Gaussian} =  \frac{1}{\sqrt{2 \, \pi} \, \sigma} \, e^{-\frac{\Delta^2}{ 2 \,\sigma^2}}  
\end{equation}
where $\sigma$ is the standard deviation.
For sake of clarity, we generically denote with $\Delta$ the difference computed for each probe with its own quantities: $\Delta \mu_{\mathrm{SNe \, Ia}} = \mu_{obs, \mathrm{SNe \, Ia}} - \mu_{th}$ for SNe Ia, $\Delta \mu_{ \mathrm{GRBs}} = \mu_{obs, \mathrm{GRBs}} - \mu_{th}$ for  GRBs, $\Delta \mathrm{log_{10}}L_X = \mathrm{log_{10}}L_{X,obs} - \mathrm{log_{10}}L_{X,th}$ for QSOs, and $\Delta d_z = d_{z,obs} - d_{z,th}$ for BAO. As shown in Section \ref{quantities}, the computation of the theoretical quantities requires the assumption of a specific cosmological model, hence we here assume a flat $\Lambda$CDM model with $\Omega_M = 0.3$ and $H_0 = 70 \, \mathrm{km} \, \mathrm{s}^{-1} \, \mathrm{Mpc}^{-1}$. Actually, we here stress that also some of the observed quantities do rely on the choice of the cosmological model.
Nevertheless, we test our results in relation to the assumptions for $\Omega_M$ and $H_0$. 

In addition, the computation of $\mu_{obs, \mathrm{GRBs}}$ and $\mathrm{log_{10}}L_{X,obs}$ requires to fix the values of the parameters of the 3D and RL relations, respectively. Thus, we use the values obtained from the fit of these relations in the different evolutionary cases, which are provided in  \citet{2022MNRAS.tmp.2639D} for GRBs and \citet{biasfreeQSO2022} for QSOs. We here stress that for the investigation of the Gaussianity assumptions, the two treatments of the evolution (fixed and varying) are equivalent as the evolutionary parameter $k$ of the fixed correction is obtained from the assumption of $\Omega_M=0.3$ \citep{DainottiQSO,2022MNRAS.tmp.2639D}, and also the function $k(\Omega_M)$ is computed for $\Omega_M = 0.3$, as assumed in our Gaussianity tests. Nevertheless, \citet{DainottiQSO} and \citet{2022MNRAS.tmp.2639D} have proved, for QSOs and GRBs respectively, that $k(\Omega_M)$ remains compatible within 1 $\sigma$ with $k(\Omega_M=0.3)$ also when $\Omega_M$ spans all the range of physical values. As a consequence, we do not expect differences between the cases with fixed $k$ and with $k(\Omega_M)$ even if we change the values assumed for $\Omega_M$ and $H_0$.

Our analysis of the Gaussianity assumption follows the one detailed in \citet{snelikelihood2022} for the \textit{Pantheon} and \textit{Pantheon +} SNe Ia samples. Specifically, we investigate the Anderson-Darling \citep{stephens1974edf,stephens1976asymptotic,stephens1977goodness,stephens1978goodness,stephens1979tests} and Shapiro-Wilk \citep{shapiro1965analysis,razali2011power} tests for normality, we compute the skewness and kurtosis of each $\Delta$ distribution, we apply the skewness \citep{doi:10.1080/00031305.1990.10475751} and kurtosis \citep{anscombe1983distribution} tests, and the ``skewness+kurtosis" test that is based on a combination of these two \citep{d1971omnibus,d1973tests}\footnote{All the analyses are carried out with the scipy Python package.}.  
Both the Anderson-Darling and Shapiro-Wilk statistical tests allow us to determine if the investigated data are drawn from a specific probability distribution, which in our case is the Gaussian distribution. These methods are commonly applied in different domains (see e.g. \citealt{stephens1974edf, razali2011power} in statistics, and \citealt{2022ApJS..261...25D} and \citealt{snelikelihood2022} in cosmology), due to their capability of identifying any small, regardless of how small, deviation from the Gaussianity in samples statistically large enough \citep[][]{stephens1976asymptotic, Leslie_86}. However, this property also limits their applicability. Indeed, as the sample size increases, these tests tend to reject the normality hypothesis even in presence of very small deviations \citep{yazici2007comparison}, which can be caused for example by ties generated by limited precision (i.e. the number of decimal digits). This weakness should be accounted for when dealing with large data samples. 

For this reason, we additionally include investigations on the skewness and the kurtosis of the $\Delta$ distributions. The skewness of a variable $x$ is the third standardized moment, which can be written as $\mathrm{E} \left[ ((x - \hat{x})/ \sigma)^3  \right]$ in the Fisher's definition. Here, E is the expectation operator and $\hat{x}$ the mean (or location). The skewness measures the asymmetry of a distribution about $\hat{x}$ by distinguishing extreme values in one tail compared to the other tail. The kurtosis is the fourth standardized moment, defined as  $\mathrm{E} \left[ ((x - \hat{x})/ \sigma)^4  \right]$, and it identifies extreme values in both the tails: the larger the kurtosis, the more populated the tails compared to the Gaussian tails, and vice versa.
If we consider the Fisher’s definition of kurtosis, in which 3 is subtracted from the result, a Gaussian distribution has both skewness and kurtosis equal to zero. Thus, the computation of skewness and kurtosis of $\Delta$ allows us to obtain information on the deviation of these distributions from normality.
As a further step, we apply the skewness, kurtosis, and ``skewness+kurtosis" tests, that determine if the values of skewness and kurtosis guarantee a statistically good Gaussian approximation. The application of all these different and complementary methods is crucial to overcome the limits of the Anderson-Darling and Shapiro-Wilk tests. It also leverages the advantages of each approach, and to obtain reliable results on the investigation of the Gaussianity assumption on $\Delta$.

\subsection{New likelihoods and fit of flat $\Lambda$CDM model}

\label{newfits}

As detailed in \citet{snelikelihood2022}, neither the \textit{Pantheon} nor the \textit{Pantheon +} SNe Ia sample passes the normality tests described in Section \ref{normalitytests}. Furthermore, we prove in Section \ref{normalresults} that also QSOs and BAO do not pass these tests. Hence, only GRBs obey the Gaussianity requisite.
As a consequence, we further extend our analysis on the non-Gaussianity of QSOs and BAO with a deeper investigation, following the same procedure employed in \citet{snelikelihood2022} for SNe Ia. As a first step, we fit the histogram of the corresponding $\Delta$ to find the best-fit distribution. To this end, we use both Python 3 \citep{van1995python} and Wolfram Mathematica 12.3 \citep{Mathematica} built-in functions. The discovered actual underlying distributions are then used in the likelihood of the cosmological analysis in place of $\cal L$$_{Gaussian}$. As we fit jointly all the probes, we choose for each one the proper likelihood and we multiply them to construct the final new likelihood ($\cal L$$_{new}$).  

Specifically, we fit with the D'Agostini method a flat $\Lambda$CDM model considering the most general case in which both $\Omega_M$ and $H_0$ are free parameters. The fit is performed in two separate cases: with the standard $\mathcal{L}_{Gaussian}$ for each probe and with the novel $\mathcal{L}_{new}$. To ensure that the MCMC process explores all possible physical regions of the $(\Omega_M, H_0)$ parameter space, we assume the wide uniform priors $0 \leq \Omega_M \leq 1$ and $50 \, \mathrm{km} \, \mathrm{s}^{-1} \, \mathrm{Mpc}^{-1} \leq H_0 \leq 100 \, \mathrm{km} \, \mathrm{s}^{-1} \, \mathrm{Mpc}^{-1}$. Additionally, broad uniform priors are considered also for the parameters of the 3D and RL relations: $-2 < a < 0$, $0 < b < 2$, $0< sv < 2$, $0.1 < g_1 < 1$, $2< n_1 < 20$, and $0 < sv_1 < 2$. 
As anticipated, when also the evolutionary parameters $k$ are free parameters of the fit, we need to assume Gaussian priors on them, which are detailed in Section \ref{EP method}.
As stressed in \citet{snelikelihood2022}, the comparison between the results obtained with $\cal L$$_{Gaussian}$ and $\cal L$$_{new}$ is pivotal to test how the change in the likelihood affects the cosmological results. Indeed, it is particularly interesting to investigate if and to what extent $\cal L$$_{new}$ changes the values of the cosmological parameters and if it performs better, even reducing the uncertainties of the cosmological parameters.

To statistically interpret our results on the $H_0$ values, we compute for each of them the z-scores $\zeta$ \citep{2022MNRAS.tmp.2639D,biasfreeQSO2022,snelikelihood2022} with respect to three $H_0$ fiducial values. By definition, $\zeta= |H_{0,i} - H_{0,our}|/ \sqrt{\sigma^2_{H_{0,i}} + \sigma^2_{H_{0,our}}}$ where $H_{0,i}$ and $\sigma_{H_{0,i}}$ are the reference value and its 1 $\sigma$ uncertainty, respectively, while $H_{0,our}$ and $\sigma_{H_{0,our}}$ are our $H_0$ value and its 1 $\sigma$ uncertainty, respectively.
The parameter $\zeta$ quantifies the deviation of the $H_0$ values obtained in this work from the fiducial ones.
The assumed reference values are:  $H_0 = 73.04 \pm 1.04 \, \mathrm{km} \, \mathrm{s}^{-1} \, \mathrm{Mpc}^{-1}$ \citep{2022ApJ...934L...7R}, $H_0 = 67.4 \pm 0.5 \, \mathrm{km} \, \mathrm{s}^{-1} \, \mathrm{Mpc}^{-1}$, \citep{planck2018}, and $H_0 = 70.00 \pm 0.13 \, \mathrm{km} \, \mathrm{s}^{-1} \, \mathrm{Mpc}^{-1}$ obtained in \citet{snelikelihood2022} when fitting a flat $\Lambda$CDM model with $\Omega_M = 0.3$, as in \citet{scolnic2018}. Furthermore, $H_0 = 70 \, \mathrm{km} \, \mathrm{s}^{-1} \, \mathrm{Mpc}^{-1}$ is also justified by the results of several works that use \textit{Pantheon} SNe Ia with other cosmological probes \citep[e.g.][]{2020JCAP...07..045D,2021MNRAS.504..300C,2022JHEAp..34...49A}.
In our notation, we use $\zeta_{P+}$,  $\zeta_{CMB}$, $\zeta_P$ when referring to the $H_0$ value derived from \textit{Pantheon +}, from the CMB, and from $H_0 = 70.00 \pm 0.13 \, \mathrm{km} \, \mathrm{s}^{-1} \, \mathrm{Mpc}^{-1}$, respectively (see Table \ref{tab:bestfit}).


\section{Results}
\label{results}

\begin{figure*}
    \centering
    \includegraphics[width=6.1cm]{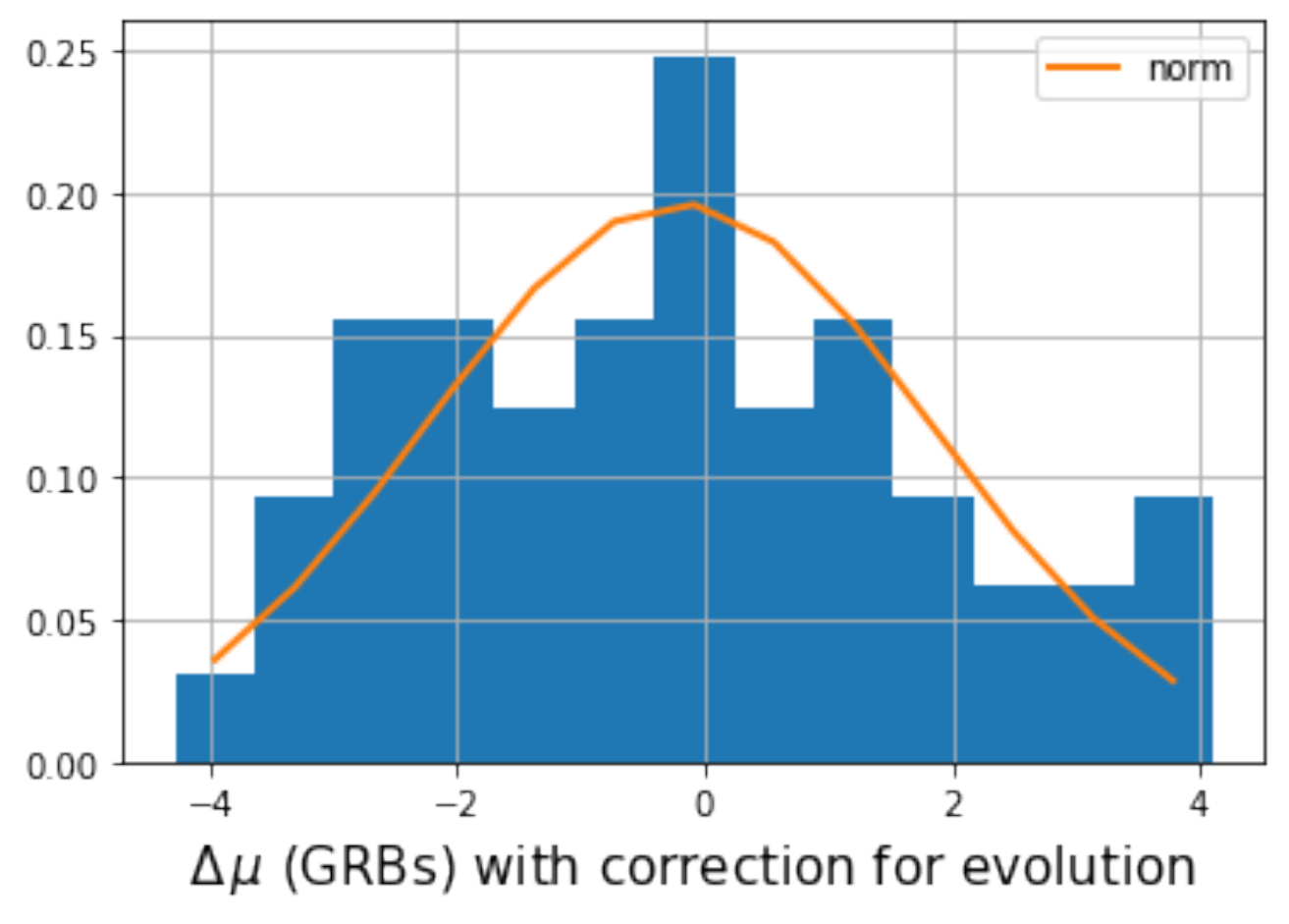}\includegraphics[width=6.1cm]{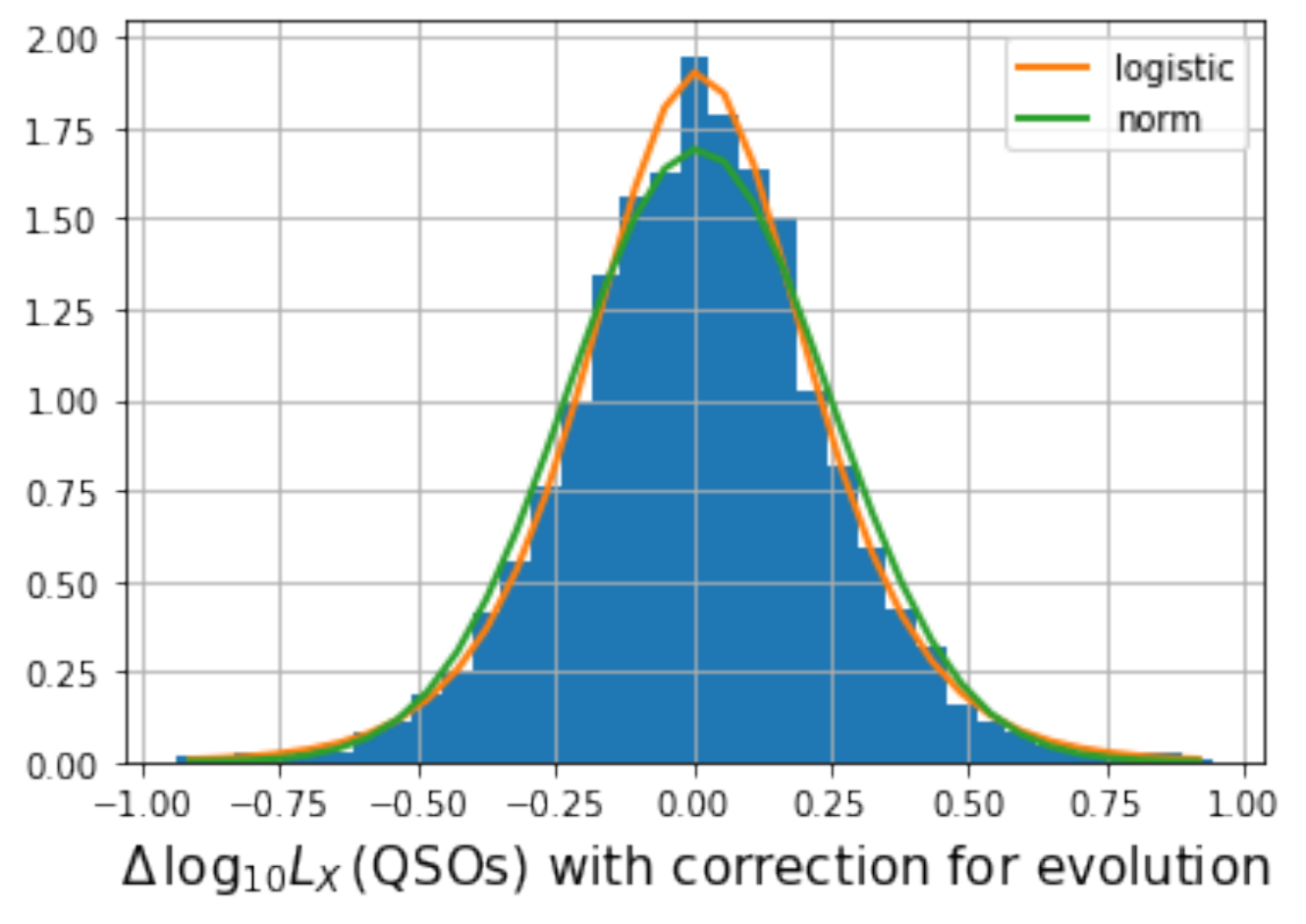}\includegraphics[width=6.1cm]{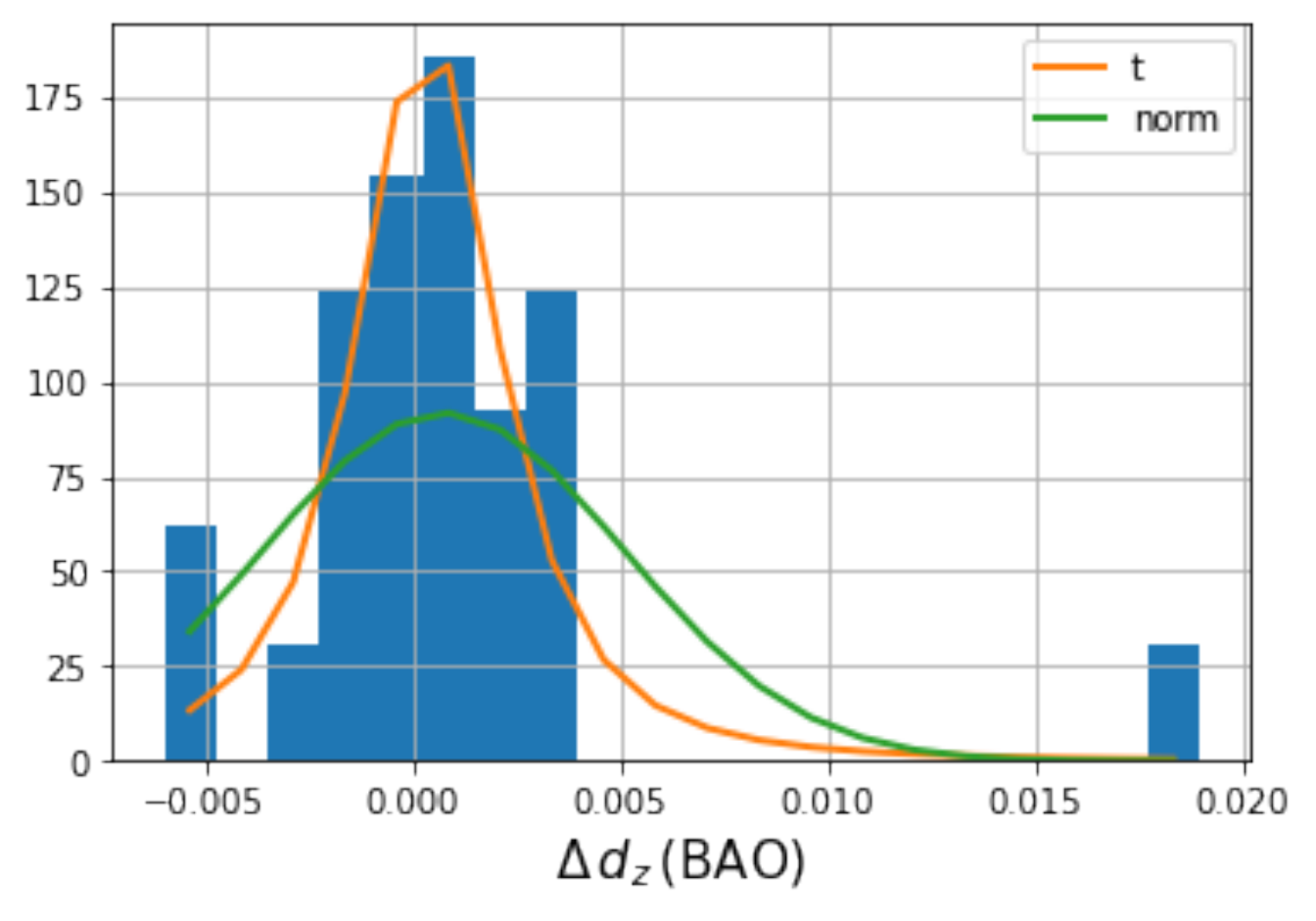}
    \caption{Normalized $\Delta$ histogram for the 50 GRBs (left panel) corrected for the redshift evolution, the 2421 QSOs corrected for the redshift evolution of luminosities (middle panel), and the 26 BAO (right panel). In the left panel, the orange curve is the best-fit distribution, which is Gaussian. In the middle and right panels, the green curve is the Gaussian distribution with the parameters that best reproduce the data, while the orange curves are the best-fit logistic and Student's t distributions, respectively.} 
    \label{fig:hist}
\end{figure*}

\subsection{Results of the Gaussianity tests}
\label{normalresults}

\citet{snelikelihood2022} have shown that both \textit{Pantheon} and \textit{Pantheon +} $\Delta \mu_{\mathrm{SNe \, Ia}}$ are not normally distributed. More specifically, for both samples the null Gaussian hypothesis is rejected at a significance level $> 15 \%$ from the Anderson-Darling test, and with p-value $< 5\%$, the minimum threshold required for the acceptance of the null hypothesis, from the Shapiro-Wilk, skewness, kurtosis, and ``skewness+kurtosis" tests. The skewness and kurtosis of the $\Delta \mu_{\mathrm{SNe \, Ia}}$ distributions reported in \citet{snelikelihood2022} are -0.19 and 0.69 for \textit{Pantheon}, and -0.19 and 4.2 for \textit{Pantheon +}, respectively (see also their Fig. 1). These results have been proved not to depend on the a priori assumption of $\Omega_M = 0.3$ and $H_0 = 70  \, \mathrm{km} \, \mathrm{s}^{-1} \, \mathrm{Mpc}^{-1}$.

On the other hand, GRBs pass all the tests, independently on the treatment of the redshift evolution and the values assumed for $\Omega_M$ and $H_0$. Indeed, the distribution in the left panel of Fig. \ref{fig:hist} could appear roughly non-Gaussian, but this is only due to the binned representation of the few data points. On the contrary, it reveals to be Gaussian under our statistical investigations.
The computation of skewness and kurtosis is not affected by the assumptions on the cosmological parameters, but it is slightly impacted by the approach employed for the redshift evolution. More precisely, when not accounting for any evolution, skewness and kurtosis of the $\Delta \mu_{\mathrm{GRBs}}$ distribution are 0.34 and -0.74, respectively, with a central value shifted toward a positive value. Indeed, in this case, the best-fit Gaussian distribution has $\hat{x} = 1.38$ and $\sigma = 2.0$. When including the correction for the evolution, 
the distribution becomes more symmetric around 0 with a skewness of 0.25, kurtosis of -0.65, and best-fit Gaussian parameters $\hat{x} =-0.21$ and $\sigma = 2.04$ (see left panel of Fig. \ref{fig:hist}).

The results of the normality tests on QSOs do not depend neither on the fixed values of $\Omega_M$ and $H_0$ nor on the approach for the redshift evolution. In all cases, the null hypothesis of Gaussianity of QSO $\Delta \mathrm{log_{10}}L_X$ is rejected by the Anderson-Darling, Shapiro-Wilk, kurtosis, and ``skewness+kurtosis" tests. Only the skewness test is fulfilled as a consequence of the value of the skewness being very close to 0. Indeed, without the correction for the evolution, the skewness is 0.07 (and the kurtosis is 0.84), hence the test is fulfilled with p-value = 0.14. Correcting for the evolution, skewness and kurtosis become 0.0007 and 0.79, respectively, and the p-value of the skewness test significantly increases up to 0.99. For a direct visualization, we show in the middle panel of Fig. \ref{fig:hist} the histogram obtained including the redshift evolution correction.
It is indeed interesting to  note from the computed skewness values and the histogram in Fig. \ref{fig:hist}, that the QSO $\Delta \mathrm{log_{10}}L_X$ distribution is symmetric around 0 and this symmetry increases when accounting for the redshift evolution. As a matter of fact, the best-fit Gaussian distribution (green curve in the middle panel of Fig. \ref{fig:hist}) has $\hat{x} = 0.009$ and $\sigma = 0.24$. Nevertheless, due to the high kurtosis, the combination of both features, considered in the ``skewness+kurtosis" test, confirms the non-Gaussianity distribution of $\Delta \mathrm{log_{10}}L_X$. 

BAO do pass none of the normality tests applied in this work, independently on the values assumed a-priori for $\Omega_M$ and $H_0$ to compute $d_{z,th}$. However, in this case, differently from GRBs and QSOs, the values of skewness and kurtosis do depend on the cosmological model assumed. This can be ascribed to the fact that the assumptions on $\Omega_M$ and $H_0$ impact the computation of $dz_{th}$ (see Eqs. \ref{rs} and \ref{DVBAO}), but not $d_{z,obs}$, hence this choice affects the difference $\Delta d_z$.
\citet{2016JCAP...06..023S} proved that, fixing $r_s(z_d) \, \cdot h = 104.57$ Mpc, as assumed in our data \citep{2018arXiv180707323S}, the best-fit values within a $\Lambda$CDM model are $\Omega_M = 0.278$ and $H_0 = 70.07  \, \mathrm{km} \, \mathrm{s}^{-1} \, \mathrm{Mpc}^{-1}$. Thus, we use these values to compute skewness and kurtosis obtaining a skewness of 2.55 and a kurtosis of 9.23 (see right panel of Fig. \ref{fig:hist}). Despite its asymmetry, the distribution is centered on 0. Indeed, the best-fit Gaussian curve (in green in the right panel of Fig. \ref{fig:hist}) has $\hat{x} = 0.001$ and $\sigma = 0.004$.

\subsection{The best-fit distributions}
\label{bestfitlikelihood}

As proved in Section \ref{normalresults}, only GRBs, independent of the treatment of the redshift evolution, can be well approximated by the Gaussian probability distribution function (PDF)
\begin{equation}
\label{gaussianlf}
\displaystyle
\mathrm{PDF_{Gaussian}} = \frac{1}{\sqrt{2 \, \pi} \, \sigma} \, e^{-\frac{1}{2} \left(\frac{x- \hat{x}}{\sigma}\right)^2}
\end{equation}
where, in our case, $x = \Delta \mu_{ \mathrm{GRBs}}$. 
Investigating \textit{Pantheon} and \textit{Pantheon +} SNe Ia, \citet{snelikelihood2022} have determined that the best-fit $\Delta \mu_{ \mathrm{SNe \, Ia}}$ distributions are a logistic and a Student's t distribution, respectively (see their Fig. 1). The logistic PDF reads as follows:
\begin{equation}
\label{logistic}
\mathrm{PDF_{logistic}} = \frac{e^{-\frac{(x-\hat{x})}{s}}}{s \, \left(1+ e^{\frac{-(x-\hat{x})}{s}}\right)^2}
\end{equation}
where $s$ is the scale parameter and the variance $\sigma^2$ is given by $\sigma^2 = (s^2 \, \pi^2)/3$. The best-fit parameters of \citet{snelikelihood2022} are $\hat{x}= -0.004$ and $s=0.08$. The generalized Student's t PDF is defined as 
\begin{equation}
\label{student}
\mathrm{PDF_{student}} = \frac{\Gamma\left(\frac{\nu +1}{2}\right)}{\sqrt{\nu \, \pi} \, s \, \Gamma \left(\frac{\nu}{2}\right)} \, \left[1 + \frac{((x- \hat{x})/s)^2}{\nu}\right]^{-\frac{\nu +1}{2}}
\end{equation}
where $\Gamma$ is the gamma function, $\nu$ are the degrees of freedom (>0), and the variance is $\sigma^2 = (s^2 \, \nu) / (\nu -2)$. The corresponding best-fit parameters in \citet{snelikelihood2022} are $\hat{x}= 0.1$, $s=0.12$, and $\nu = 3.8$. 

Fitting the QSO $\Delta \mathrm{log_{10}}L_X$, we find that, for all evolutionary cases, the best-fit distribution is the logistic distribution, whose PDF is the one in Eq. \eqref{logistic}. The middle panel of Fig. \ref{fig:hist} shows the normalized histogram, once accounted for the correction for evolution, superimposed with the best-fit logistic curve (in orange) with $\hat{x}= 0.009$ and $s=0.13$ and with the best-fit Gaussian curve (in green) with  $\hat{x}= 0.009$ and $\sigma=0.24$. In the case without evolution, the best-fit parameters for the logistic and the normal distribution are $\hat{x}= 0.0006$ and $s=0.13$, and $\hat{x}= 0.002$ and $\sigma=0.24$, respectively.

For BAO, the best-fit of the $\Delta d_z$ histogram (right panel of Fig. \ref{fig:hist}) is the Student's t distribution in Eq. \eqref{student}.
The best-fit Student's t curve shown in orange in the right panel of Fig. \ref{fig:hist} has $\hat{x} = 0.0003$, $s = 0.002$, and $\nu = 2.23$, while the best-fit Gaussian curve in green has $\hat{x} = 0.001$ and $\sigma = 0.004$. For both QSOs and BAO, the logistic and the Student's t distributions, respectively, better reproduce $\Delta$ in all the features (e.g. peak, central width, and tails) compared to the Gaussian PDF. We here stress that these investigations on the Gaussianity assumption and the proper distribution are crucial to validate the results of cosmological analyses and have to be performed on each probe used and also on future new samples.

\subsection{Cosmological results by using Gaussian and the best-fit likelihoods}
\label{cosmologicalfits}

\begin{figure*}
\centering
\begin{subfigure}[b]{.45\linewidth}
\includegraphics[width=\linewidth]{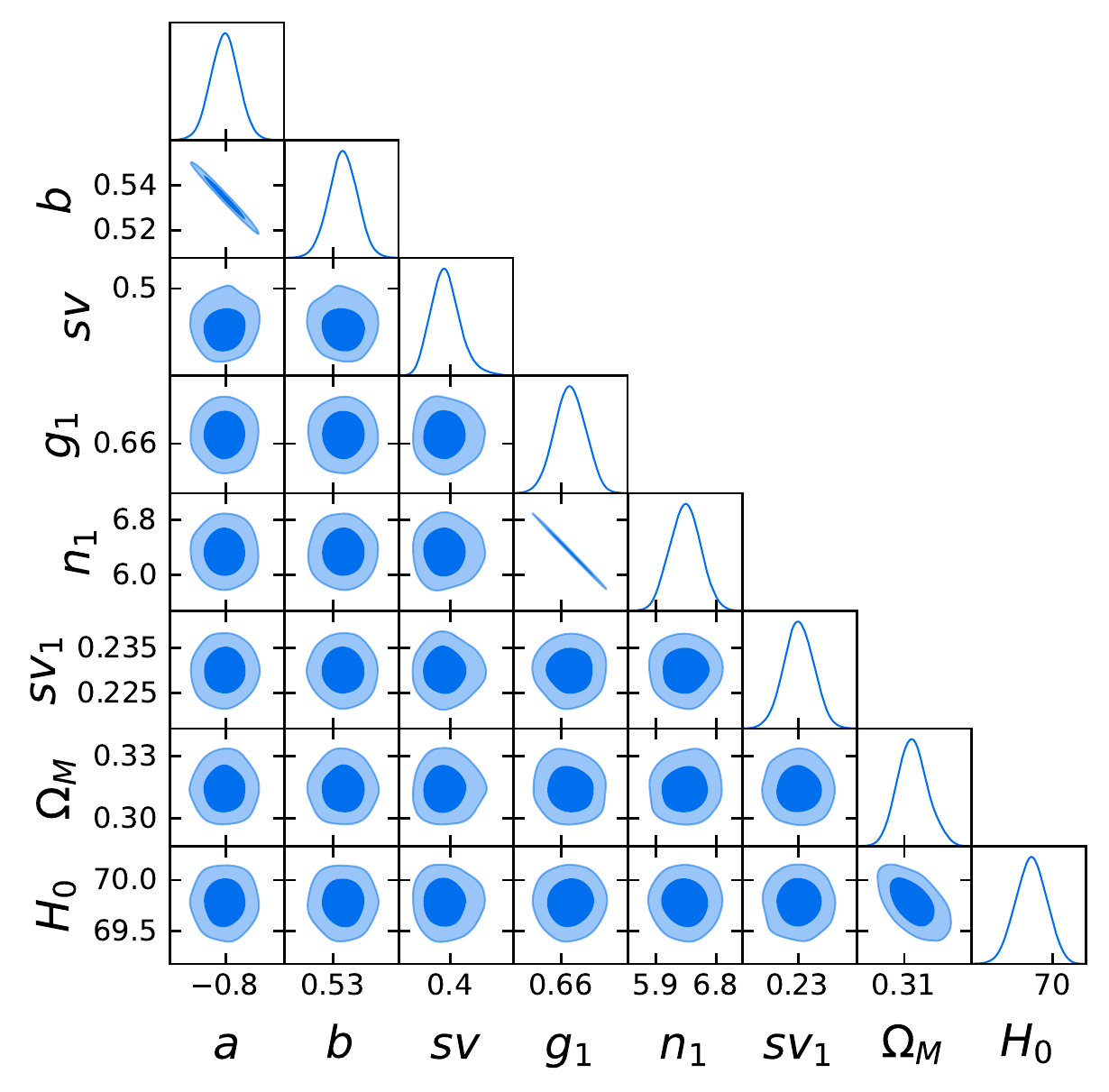}
\caption{ \textit{Pantheon} SNe Ia + GRBs + QSOs + BAO with $\cal L$$_{Gaussian}$}\label{fig: Om+H0 No Ev_P_Gauss}
\end{subfigure}
\begin{subfigure}[b]{.45\linewidth}
\includegraphics[width=\linewidth]{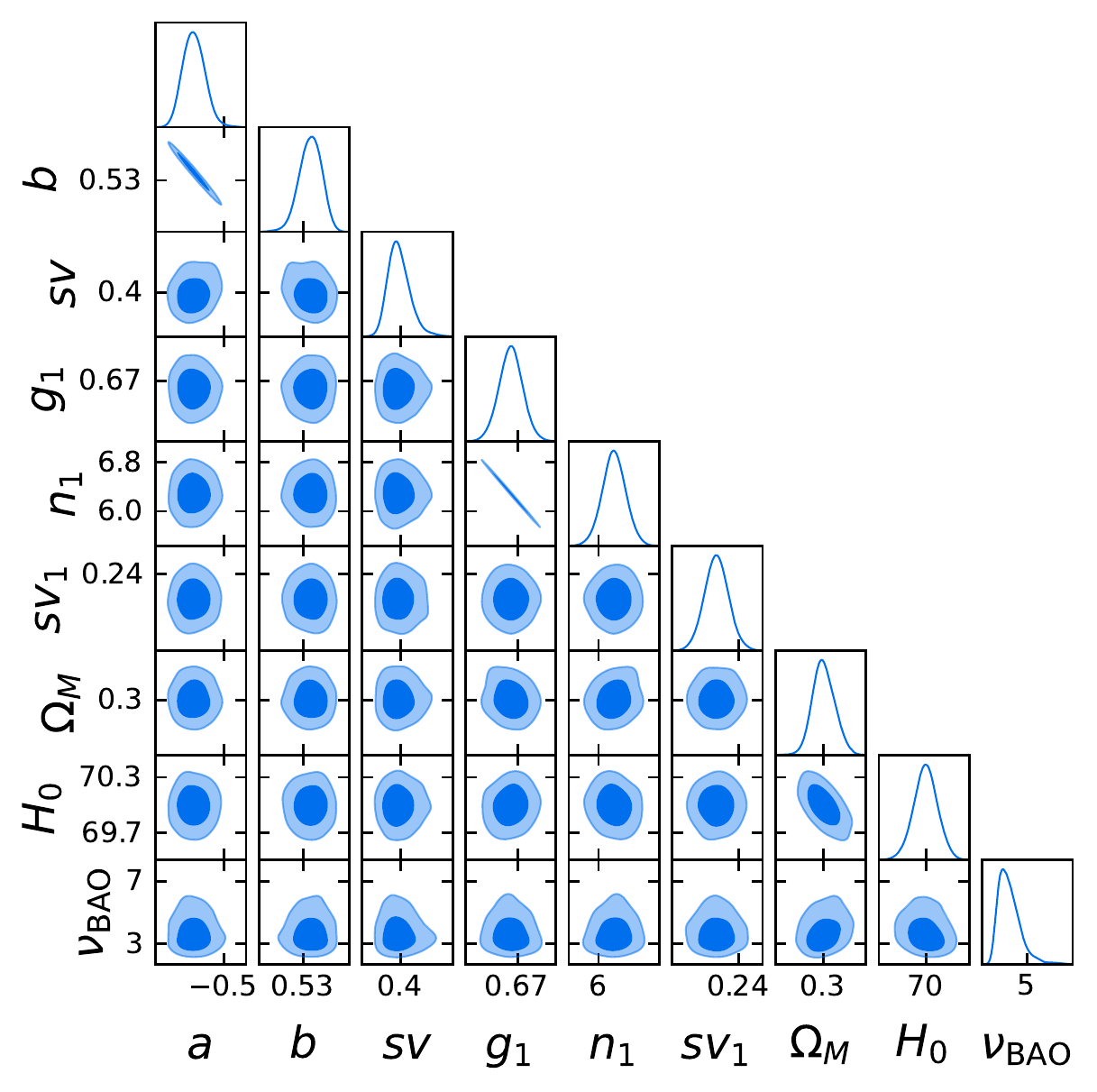}
\caption{ \textit{Pantheon} SNe Ia + GRBs + QSOs + BAO with $\cal L$$_{new}$}\label{fig: Om+H0 No Ev_P_New}
\end{subfigure}
\begin{subfigure}[b]{.45\linewidth}
\includegraphics[width=\linewidth]{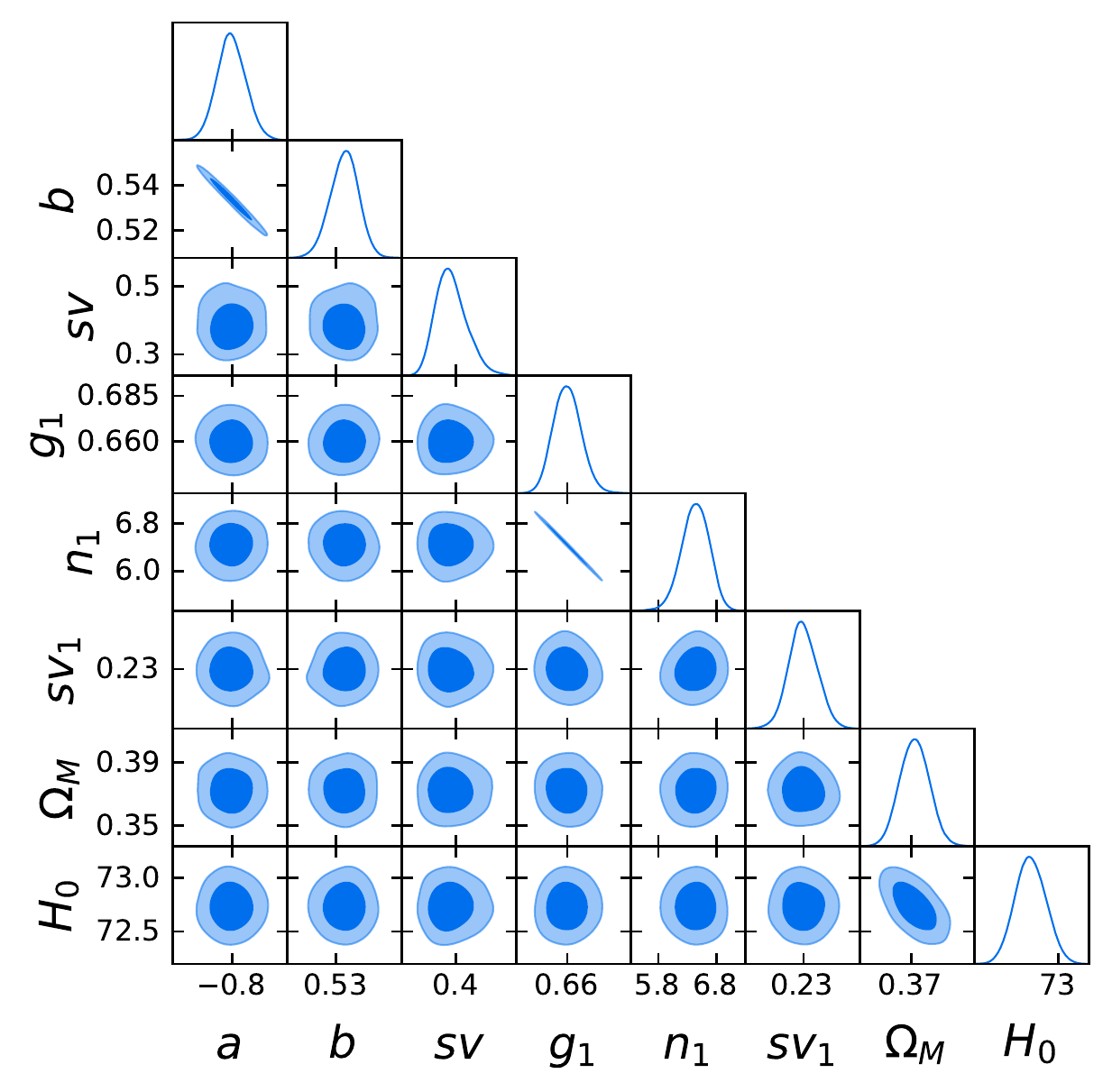}
\caption{ \textit{Pantheon +} SNe Ia + GRBs + QSOs + BAO with $\cal L$$_{Gaussian}$}\label{fig: Om+H0 No Ev_P+_Gauss}
\end{subfigure}
\begin{subfigure}[b]{.45\linewidth}
\includegraphics[width=\linewidth]{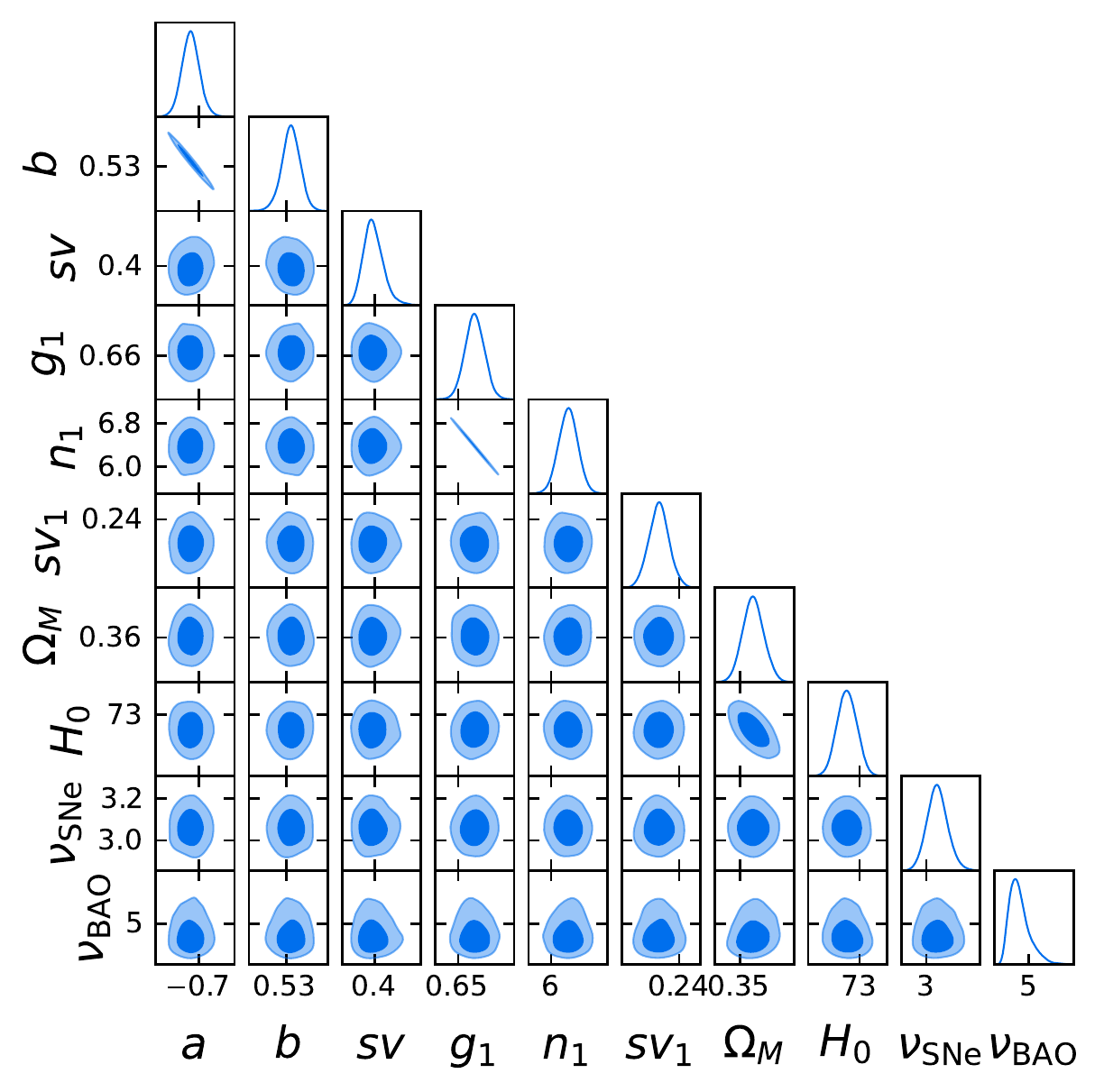}
\caption{ \textit{Pantheon +} SNe Ia + GRBs + QSOs + BAO with $\cal L$$_{new}$}\label{fig: Om+H0 No Ev_P+_New}
\end{subfigure}
\caption{Fit of the flat $\Lambda$CDM model without correction for redshift evolution.}
\label{fig: Om+H0 No Ev}
\end{figure*}

\begin{figure*}
\centering
\begin{subfigure}[b]{.45\linewidth}
\includegraphics[width=\linewidth]{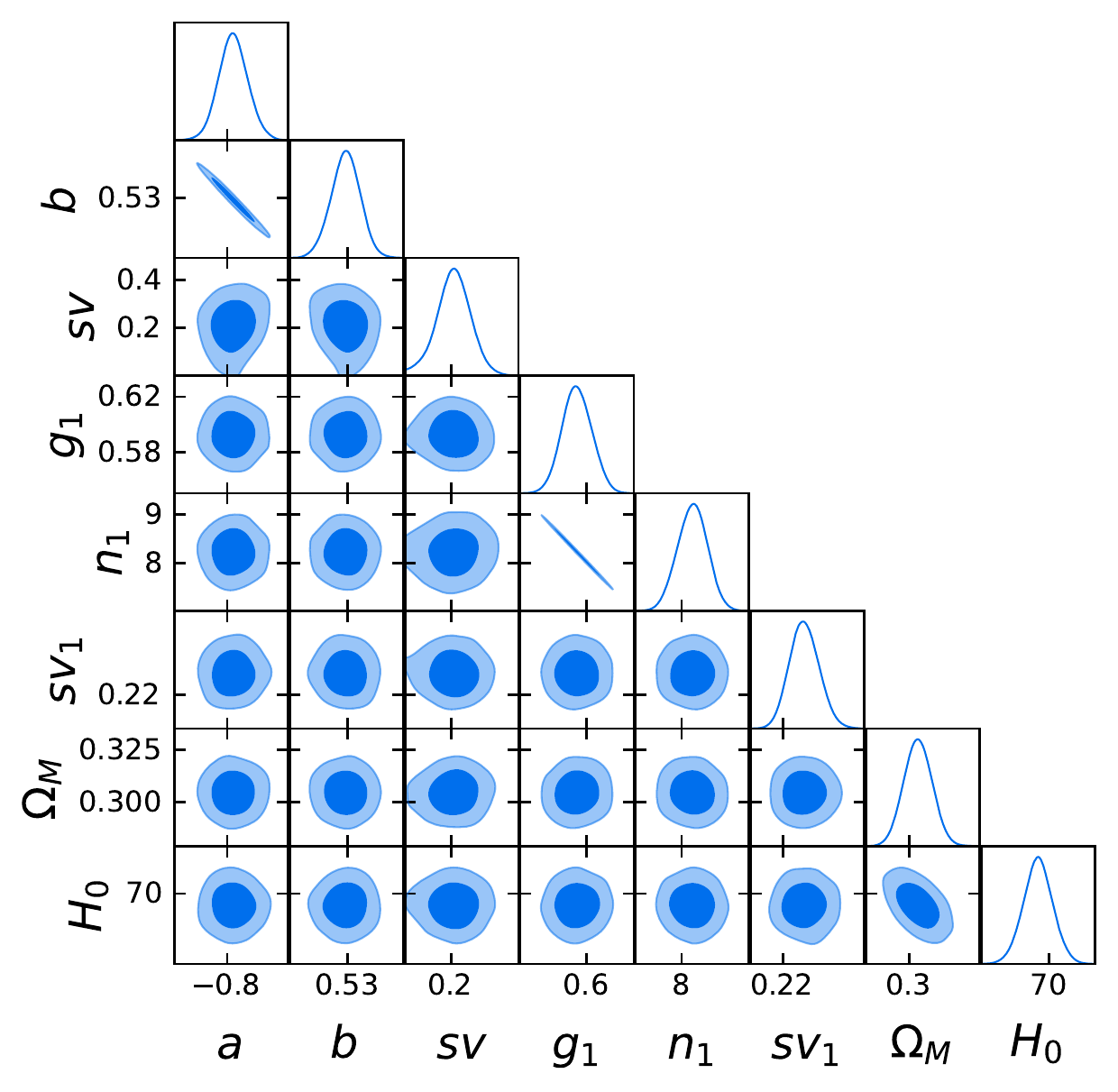}
\caption{ \textit{Pantheon} SNe Ia + GRBs + QSOs + BAO with $\cal L$$_{Gaussian}$}\label{fig: Om+H0 fixed Ev_P_Gauss}
\end{subfigure}
\begin{subfigure}[b]{.45\linewidth}
\includegraphics[width=\linewidth]{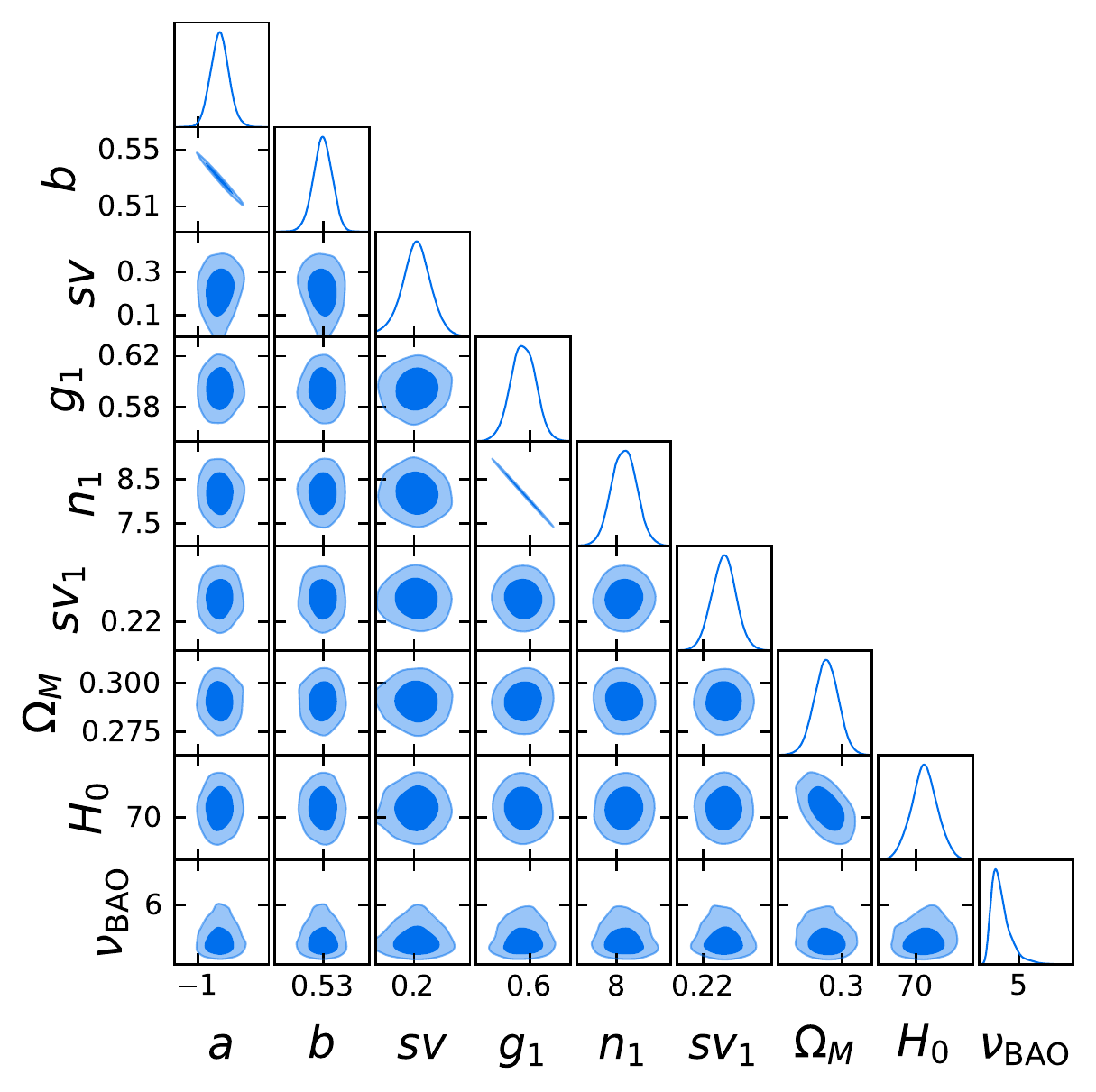}
\caption{ \textit{Pantheon} SNe Ia + GRBs + QSOs + BAO with $\cal L$$_{new}$}\label{fig: Om+H0 fixed Ev_P_New}
\end{subfigure}
\begin{subfigure}[b]{.45\linewidth}
\includegraphics[width=\linewidth]{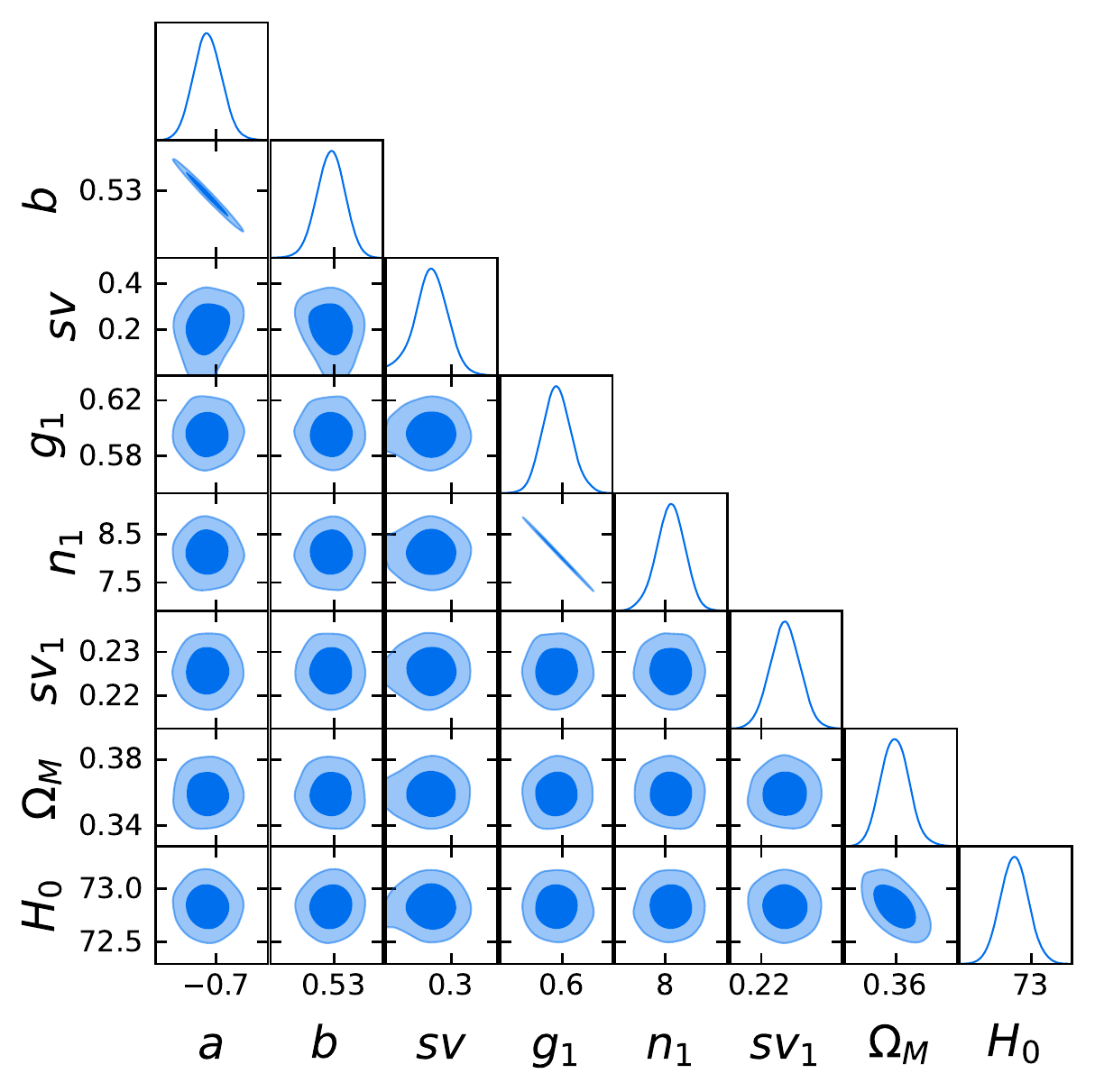}
\caption{ \textit{Pantheon +} SNe Ia + GRBs + QSOs + BAO with $\cal L$$_{Gaussian}$}\label{fig: Om+H0 fixed Ev_P+_Gauss}
\end{subfigure}
\begin{subfigure}[b]{.45\linewidth}
\includegraphics[width=\linewidth]{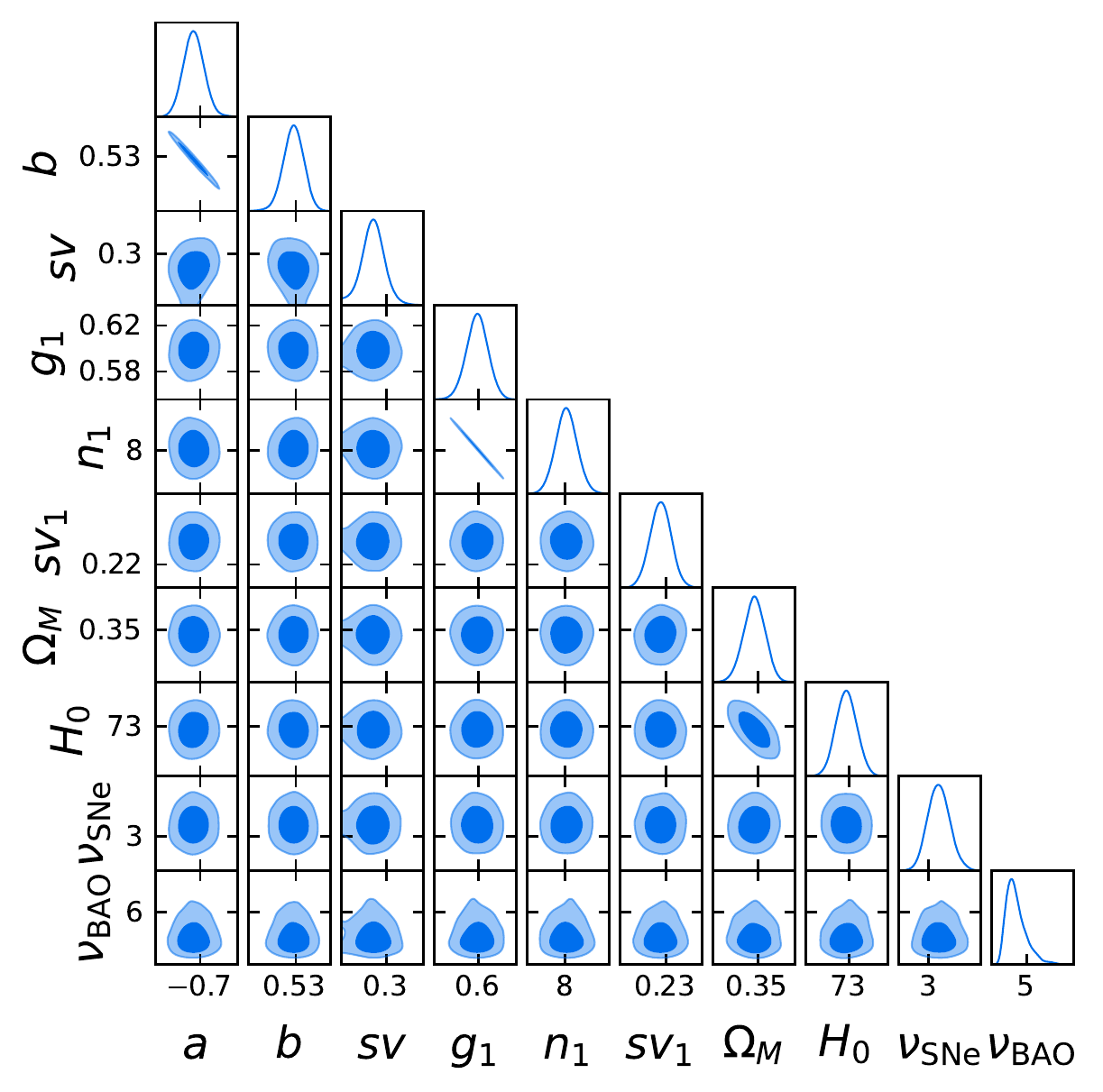}
\caption{ \textit{Pantheon +} SNe Ia + GRBs + QSOs + BAO with $\cal L$$_{new}$}\label{fig: Om+H0 fixed Ev_P+_New}
\end{subfigure}
\caption{Fit of the flat $\Lambda$CDM model with fixed correction for redshift evolution.}
\label{fig: Om+H0 fixed Ev}
\end{figure*}

\begin{figure*}
\centering
\begin{subfigure}[b]{.45\linewidth}
\includegraphics[width=\linewidth]{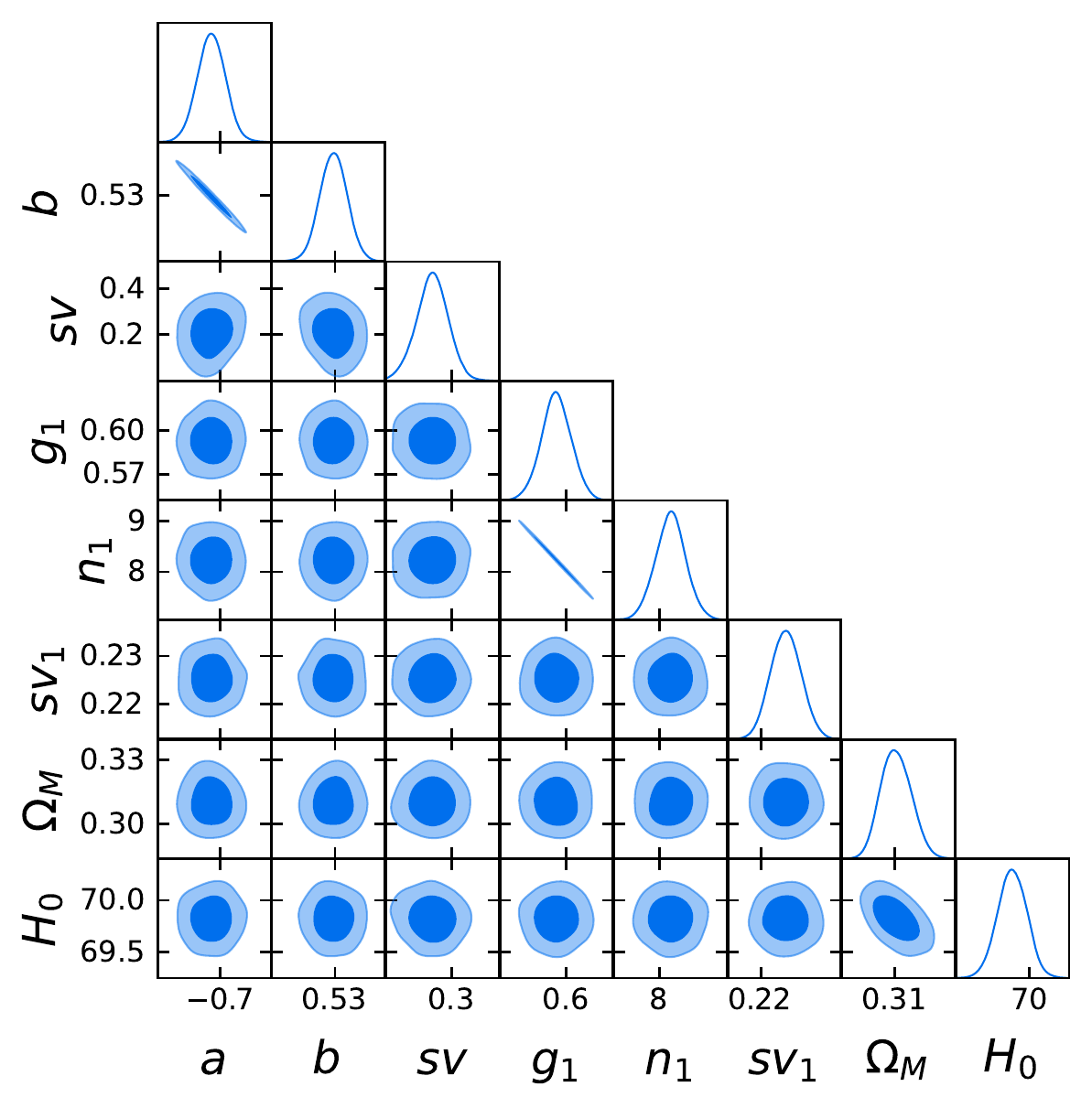}
\caption{ \textit{Pantheon} SNe Ia + GRBs + QSOs + BAO with $\cal L$$_{Gaussian}$}\label{fig: Om+H0 var Ev_P_Gauss}
\end{subfigure}
\begin{subfigure}[b]{.45\linewidth}
\includegraphics[width=\linewidth]{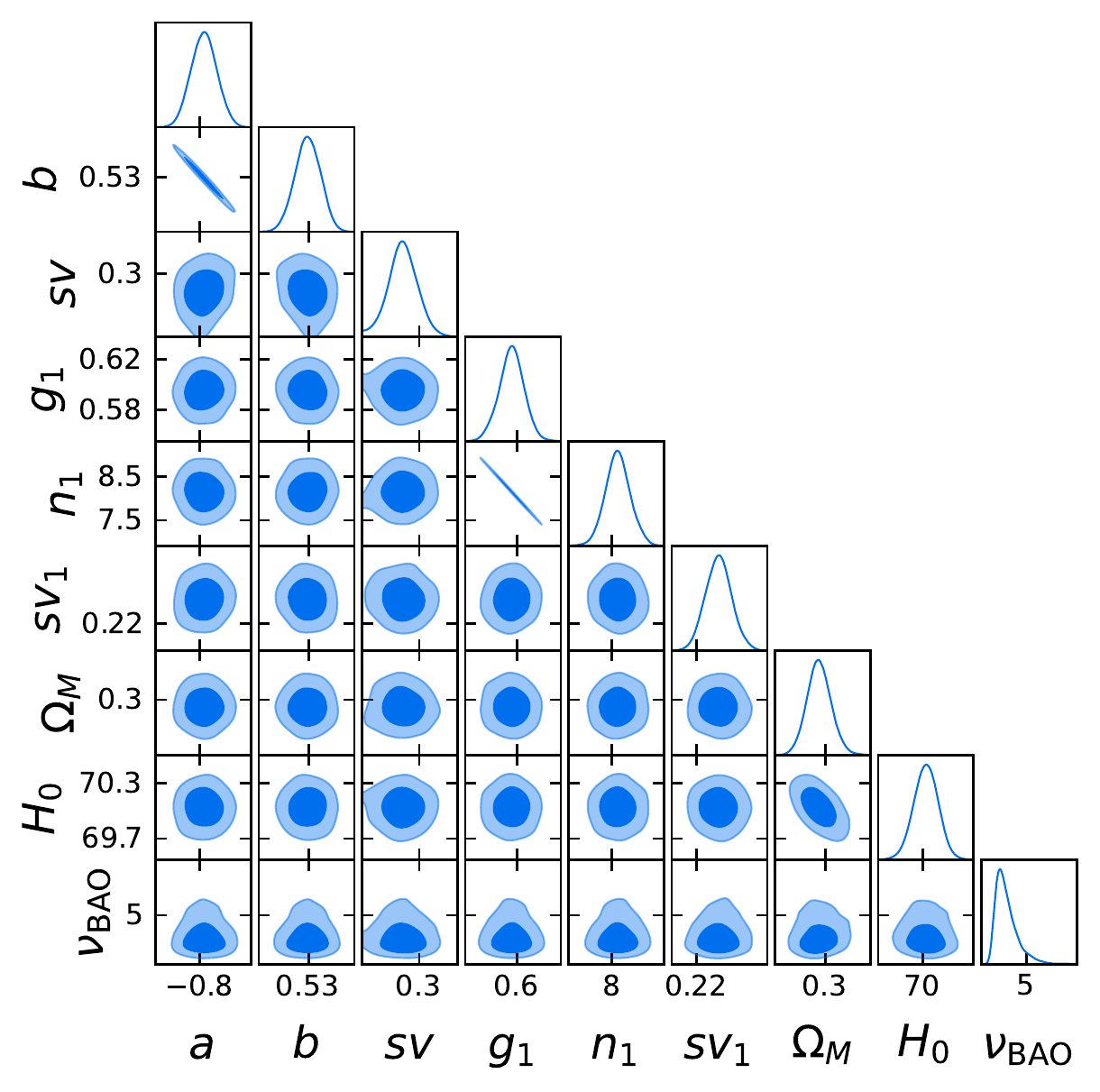}
\caption{ \textit{Pantheon} SNe Ia + GRBs + QSOs + BAO with $\cal L$$_{new}$}\label{fig: Om+H0 var Ev_P_New}
\end{subfigure}
\begin{subfigure}[b]{.45\linewidth}
\includegraphics[width=\linewidth]{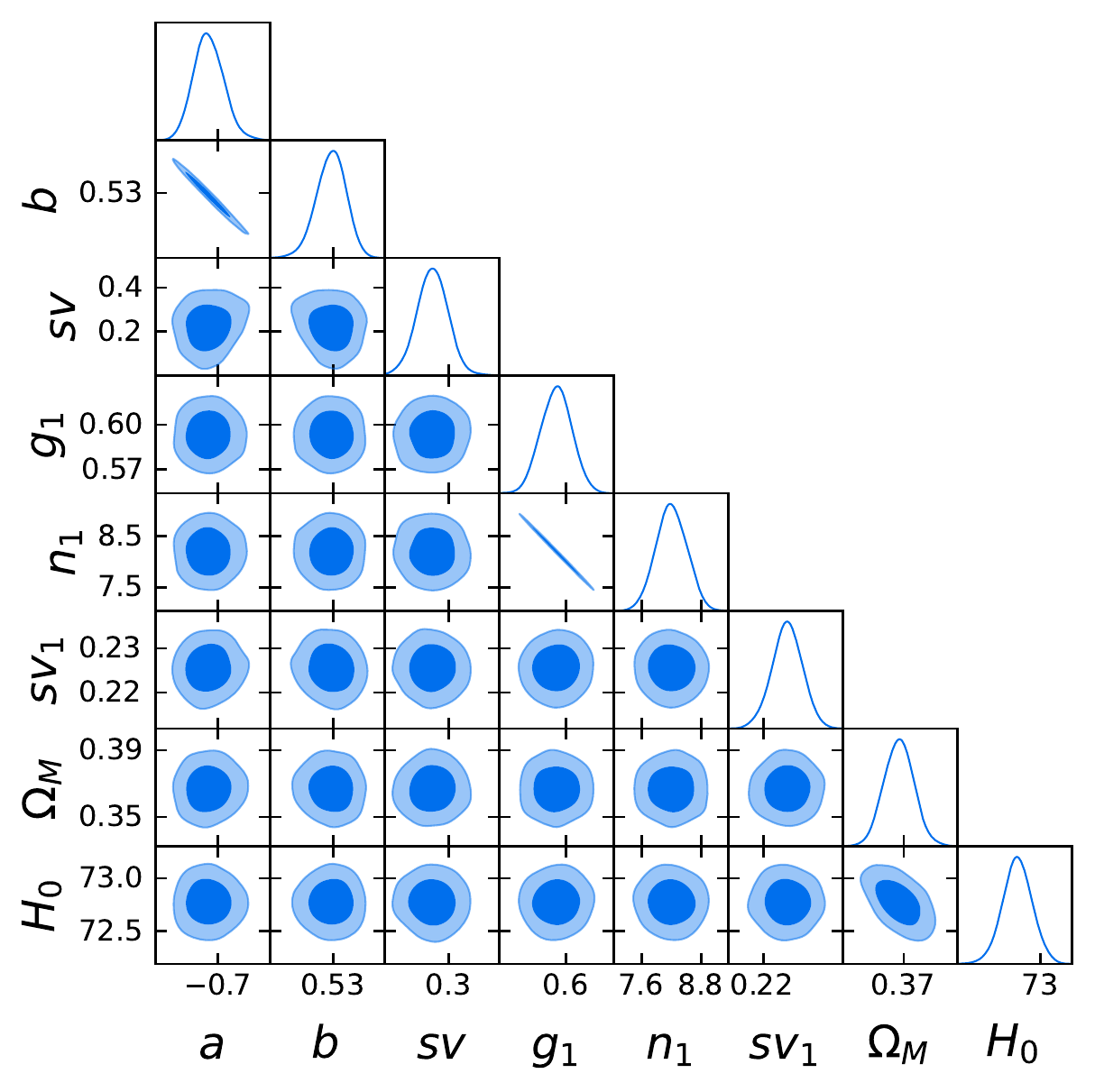}
\caption{ \textit{Pantheon +} SNe Ia + GRBs + QSOs + BAO with $\cal L$$_{Gaussian}$}\label{fig: Om+H0 var Ev_P+_Gauss}
\end{subfigure}
\begin{subfigure}[b]{.45\linewidth}
\includegraphics[width=\linewidth]{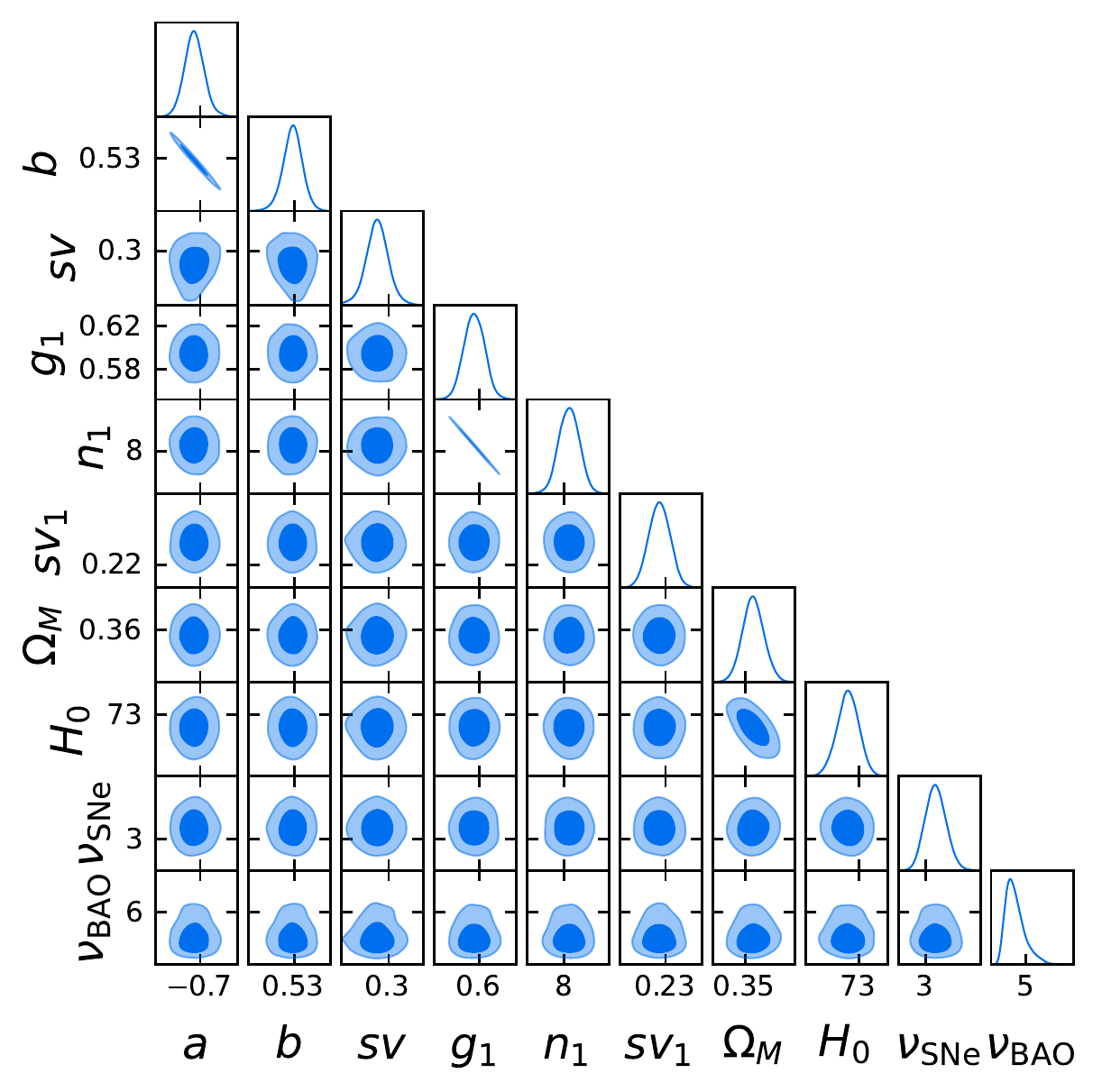}
\caption{ \textit{Pantheon +} SNe Ia + GRBs + QSOs + BAO with $\cal L$$_{new}$}\label{fig: Om+H0 var Ev_P+_New}
\end{subfigure}
\caption{Fit of the flat $\Lambda$CDM model with correction for redshift evolution as a function of cosmology.}
\label{fig: Om+H0 var Ev}
\end{figure*}

\begin{figure*}
\centering
\begin{subfigure}[b]{.49\linewidth}
\includegraphics[width=\linewidth] {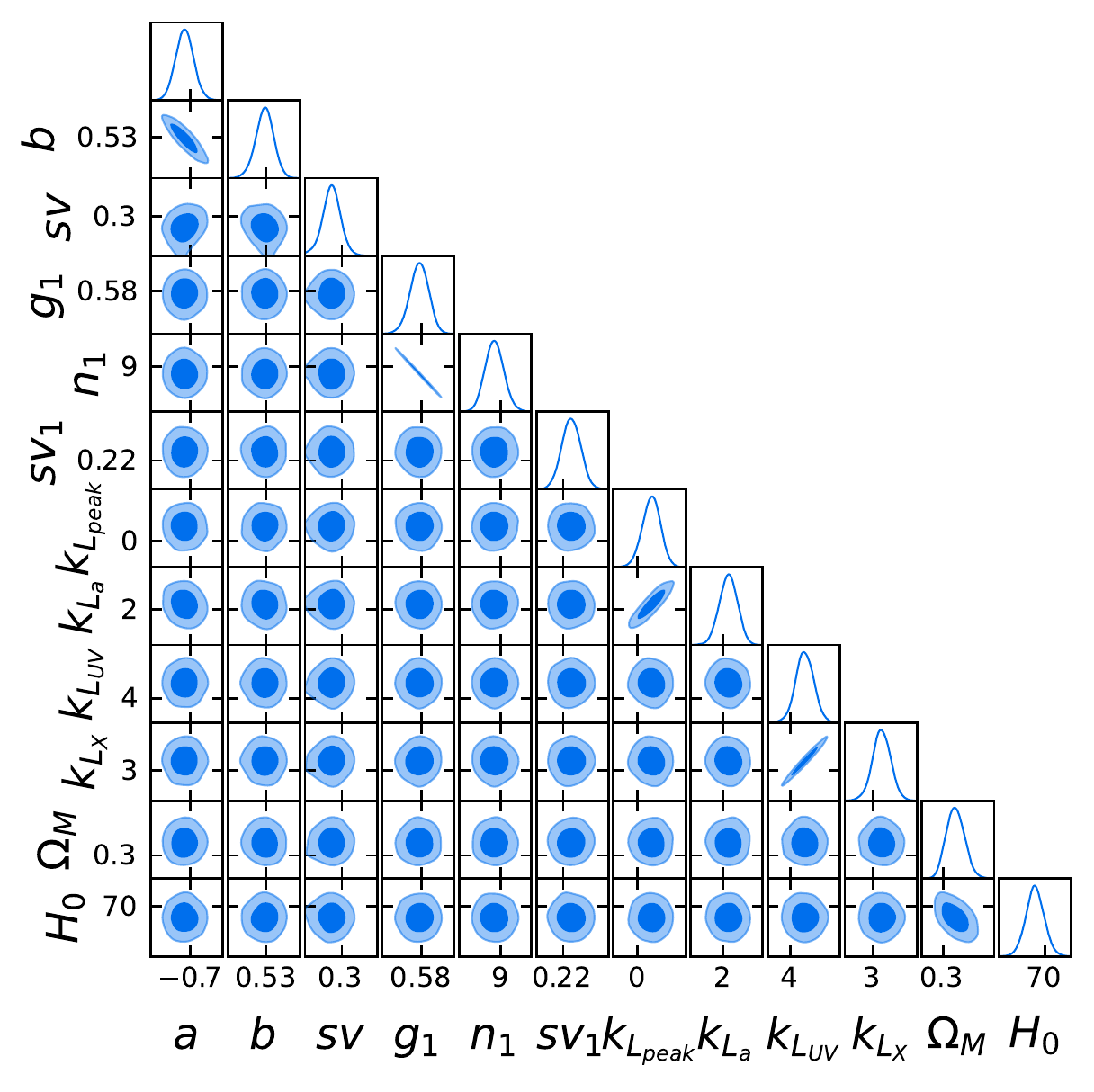}
\caption{ \textit{Pantheon} SNe Ia + GRBs + QSOs + BAO}
\end{subfigure}
\begin{subfigure}[b]{.49\linewidth}
\includegraphics[width=\linewidth]{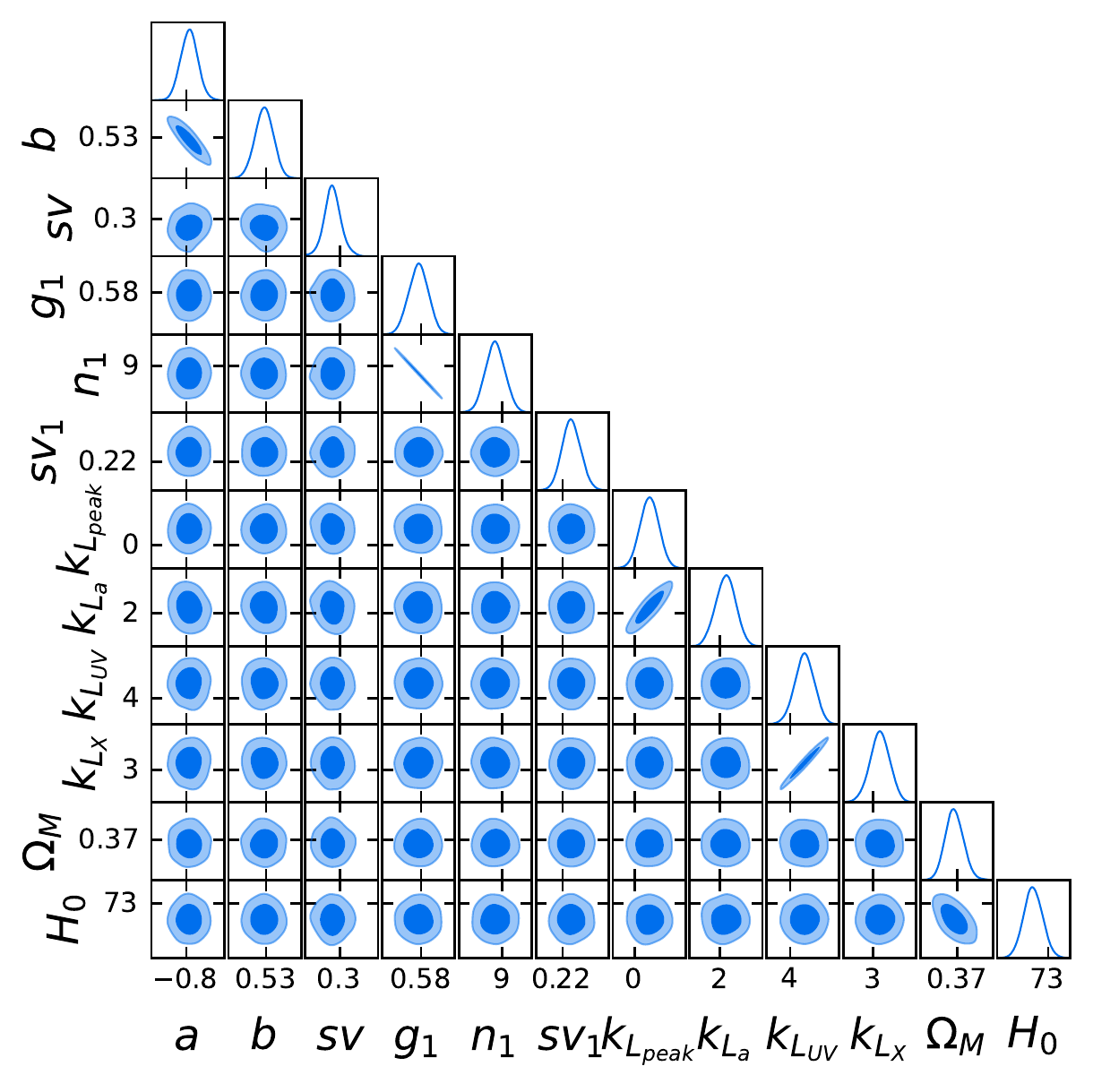}
\caption{ \textit{Pantheon +} SNe Ia + GRBs + QSOs + BAO}
\end{subfigure}
\caption{Fit of the flat $\Lambda$CDM model with $\cal L$$_{Gaussian}$ and the evolutionary parameters $k$ for QSOs and GRBs as additional free parameters. Left panel shows the case with \textit{Pantheon}, while the right panel the one with \textit{Pantheon +}.}
\label{fig: kfree Gaussian}
\end{figure*}

\begin{figure*}
\centering
\begin{subfigure}[b]{.49\linewidth}
\includegraphics[width=\linewidth]{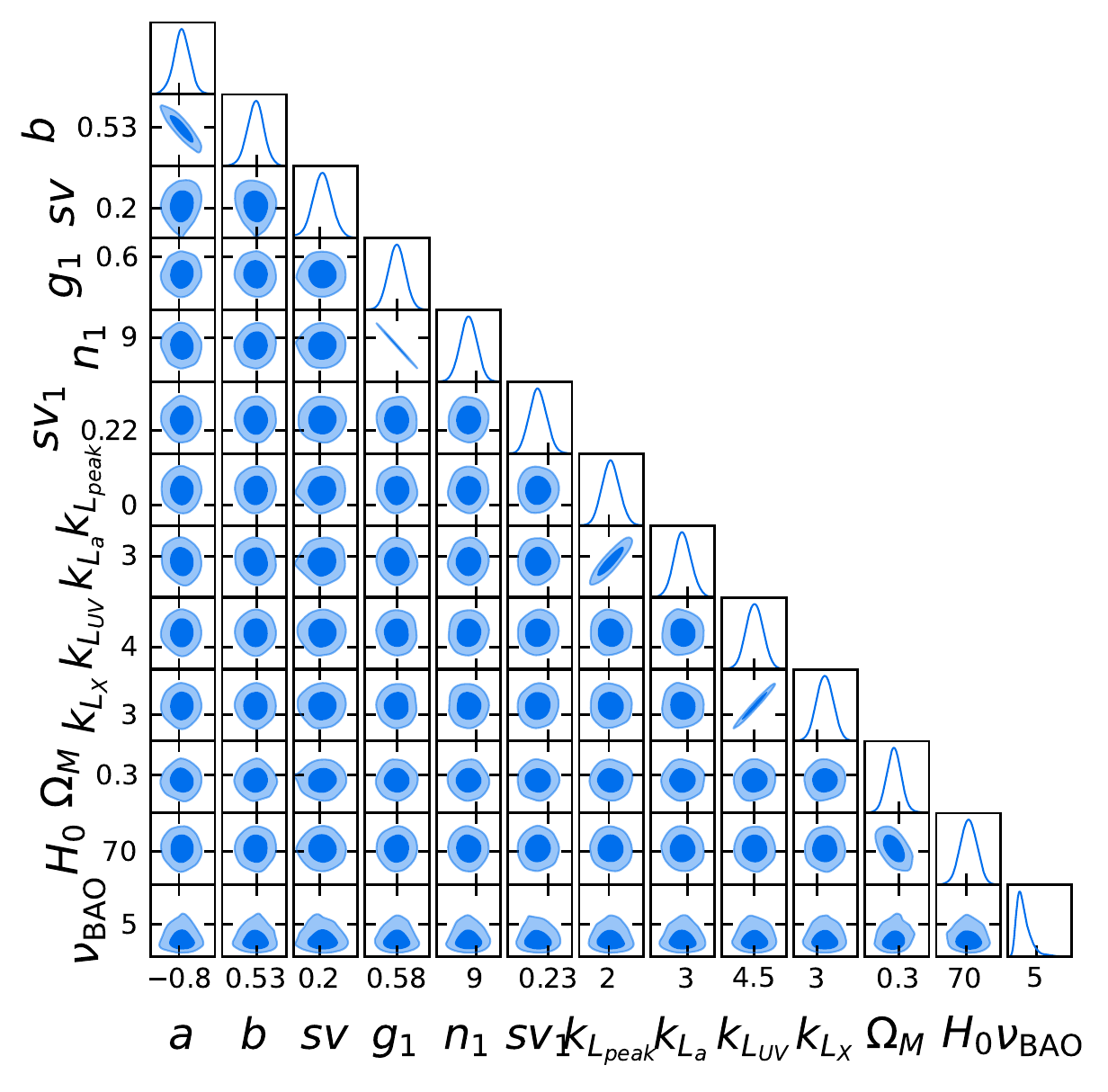}
\caption{ \textit{Pantheon} SNe Ia + GRBs + QSOs + BAO}
\end{subfigure}
\begin{subfigure}[b]{.49\linewidth}
\includegraphics[width=\linewidth]{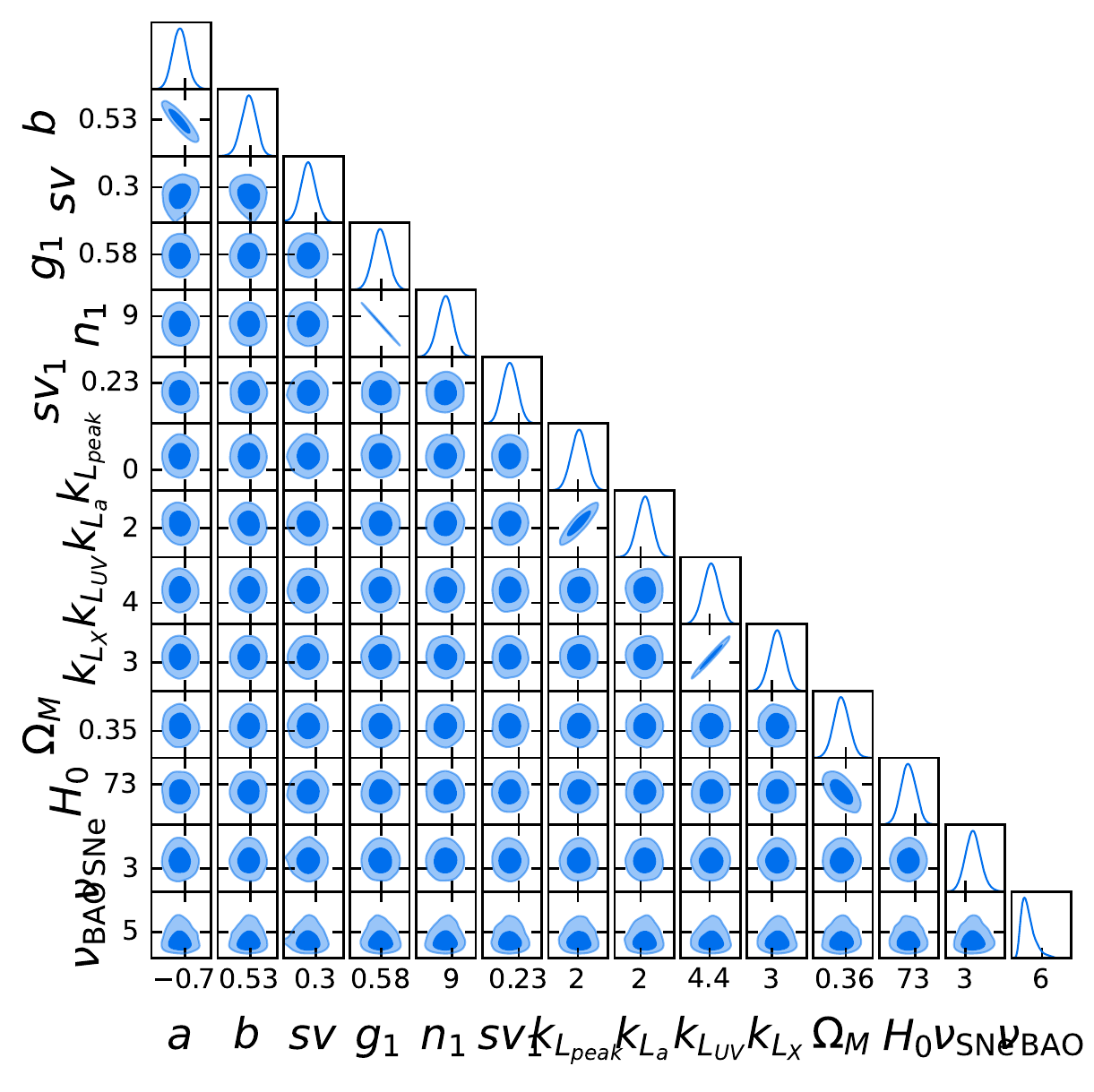}
\caption{ \textit{Pantheon +} SNe Ia + GRBs + QSOs + BAO}
\end{subfigure}
\caption{Fit of the flat $\Lambda$CDM model with $\cal L$$_{New}$ and the evolutionary parameters $k$ for QSOs and GRBs as additional free parameters. Left panel shows the case with \textit{Pantheon}, while the right panel the one with \textit{Pantheon +}.}
\label{fig: kfree newlikelihoods}
\end{figure*}

As detailed in Section \ref{newfits}, we fit a flat $\Lambda$CDM model with $\Omega_M$ and $H_0$ free parameters using the joint samples of SNe Ia (from \textit{Pantheon}) + GRBs + QSOs + BAO and SNe Ia (from \textit{Pantheon +}) + GRBs + QSOs + BAO both considering $\mathcal{L}_{Gaussian}$ for all the probes and using $\mathcal{L}_{new}$.
As the Student's t distribution depends also on the parameter $\nu$, in $\mathcal{L}_{new}$ we consider it as an additional free parameter for \textit{Pantheon +} SNe Ia ($\nu_{\mathrm{SNe}}$) and BAO ($\nu_{\mathrm{BAO}}$), imposing the wide uniform prior $0 < \nu < 10$. All results are reported in Table \ref{tab:bestfit} and Figs. \ref{fig: Om+H0 No Ev}, \ref{fig: Om+H0 fixed Ev}, \ref{fig: Om+H0 var Ev}, \ref{fig: kfree Gaussian}, and \ref{fig: kfree newlikelihoods} according to the different treatments of redshift evolution for GRBs and QSOs. For easier and more direct comparison, we also include in Table \ref{tab:bestfit} the results obtained in \citet{snelikelihood2022} with only \textit{Pantheon} and \textit{Pantheon +} SNe Ia samples. In these cases, $\cal L$$_{new}$ corresponds to the likelihood with the logistic and the Student's t distribution, respectively for \textit{Pantheon} and \textit{Pantheon +}. 
We here discuss our main findings.

\begin{table*}
\caption{Best-fit $\Omega_M$ and $H_0$ values with 1 $\sigma$ uncertainty for GRBs+QSOs+BAO together with \textit{Pantheon} or \textit{Pantheon +} SNe Ia, as specified in the left column, for all evolutionary cases and both likelihoods $\cal L$$_{new}$ and $\cal L$$_{Gaussian}$ described in this work. For easier comparison, we also report the results from only \textit{Pantheon} and \textit{Pantheon +} samples as obtained in \citet{snelikelihood2022}. In the corresponding part of the table, $\cal L$$_{new}$ is the likelihood with the logistic and the Student's t distribution, respectively for \textit{Pantheon} and \textit{Pantheon +}. $\zeta_{CMB}$, $\zeta_{P}$, and $\zeta_{P+}$ are the $\zeta$ computed as defined in Section \ref{newfits}. $H_0$ is in units of $\mathrm{km} \, \mathrm{s}^{-1} \, \mathrm{Mpc}^{-1}$.}
\begin{centering}
\begin{adjustbox}{width=\textwidth,center}
\begin{tabular}{ccccccccccc}
\hline
\multicolumn{6}{c}{$\cal L$$_{new}$}&\multicolumn{5}{c}{$\cal L$$_{Gaussian}$}\tabularnewline
\hline
\hline
\textit{Pantheon} \citep{snelikelihood2022} & $H_0$ & $\Omega_M$ & $\zeta_{CMB}$ & $\zeta_{P}$ & $\zeta_{P+}$ & $H_0$ & $\Omega_M$ & $\zeta_{CMB}$ & $\zeta_{P}$ & $\zeta_{P+}$
\tabularnewline
\hline
\hline
& $70.21 \pm 0.20$ & $0.285 \pm 0.012$ & 5.89 & 0.88 & 2.67 & $69.99 \pm 0.34$ & $0.300 \pm 0.021$ & 4.88 & 0.03 & 2.79 \tabularnewline
\hline
\hline
GRBs+QSOs+BAO+\textit{Pantheon} & $H_0$ & $\Omega_M$ & $\zeta_{CMB}$ & $\zeta_{P}$ & $\zeta_{P+}$ & $H_0$ & $\Omega_M$ & $\zeta_{CMB}$ & $\zeta_{P}$ & $\zeta_{P+}$
\tabularnewline
\hline
\hline
No Evolution & $70.01 \pm 0.14$ & $0.300 \pm 0.007$ & 5.03 & 0.05 & 2.89 & $69.78 \pm 0.15$ & $0.315 \pm 0.008$  & 5.25  & 1.11 & 3.1 \tabularnewline
\hline
Fixed Evolution & $70.09 \pm 0.14$ & $0.291 \pm 0.007$ & 5.18 & 0.47 & 2.81 & $69.88 \pm 0.15$ & $0.305 \pm 0.007$  & 5.44  & 0.60 & 3.01 \tabularnewline
\hline
Varying Evolution & $70.03 \pm 0.14$ & $0.297 \pm 0.007$ & 5.07 & 0.16 & 2.87 & $69.82 \pm 0.15$  & $0.311 \pm 0.007$  & 5.33  & 0.91 & 3.06 
\tabularnewline
\hline
\hline
\textit{Pantheon +} \citep{snelikelihood2022} & $H_0$ & $\Omega_M$ & $\zeta_{CMB}$ & $\zeta_{P}$ & $\zeta_{P+}$ & $H_0$ & $\Omega_M$ & $\zeta_{CMB}$ & $\zeta_{P}$ & $\zeta_{P+}$
\tabularnewline
\hline
\hline
& $72.93 \pm 0.16$ & $0.352 \pm 0.011$ & 11.22 & 14.21 & 0.10 & $72.85 \pm 0.24$ & $0.361 \pm 0.019$ & 10.48 & 10.44 & 0.18 \tabularnewline
\hline
\hline
GRBs+QSOs+BAO+\textit{Pantheon +} & $H_0$ & $\Omega_M$ & $\zeta_{CMB}$ & $\zeta_{P}$ & $\zeta_{P+}$ & $H_0$ & $\Omega_M$ & $\zeta_{CMB}$ & $\zeta_{P}$ & $\zeta_{P+}$
\tabularnewline
\hline
\hline
No Evolution & $72.83 \pm 0.14$ & $0.361 \pm 0.009$ & 10.46 & 14.81 & 0.20 & $72.74 \pm 0.15$ & $0.372 \pm 0.009$  & 10.23 & 13.80 & 0.68 \tabularnewline
\hline
Fixed Evolution & $72.96 \pm 0.13$ & $0.347 \pm 0.009$ & 10.76 & 16.10 & 0.08 & $72.83 \pm 0.14$ & $0.359 \pm 0.009$  & 10.46  & 14.81 & 0.20 \tabularnewline
\hline
Varying Evolution & $72.86 \pm 0.13$ & $0.356 \pm 0.009$ & 10.57 & 15.56 & 0.17 & $72.77 \pm 0.14$ & $0.367 \pm 0.009$  & 10.34  & 14.50 & 0.26 \tabularnewline
\hline
\end{tabular}
\end{adjustbox}
\label{tab:bestfit}
\par\end{centering}
\end{table*}

\begin{itemize}
    \item For both SNe Ia samples considered and independently on the evolutionary treatment employed, the best-fit values of $\Omega_M$ and $H_0$ are compatible within 1 $\sigma$ when comparing the application of $\mathcal{L}_{new}$ and $\mathcal{L}_{Gaussian}$, even if the use of $\mathcal{L}_{new}$ slightly lowers the $\Omega_M$ values (see Table \ref{tab:bestfit}). This result is completely consistent with the analysis on SNe Ia reported by \citet{snelikelihood2022}, in which the values of the cosmological parameters proved not to be affected by the choice of the likelihood function. Concerning the uncertainties on $\Omega_M$ and $H_0$, they do not show any dependence on the correction for the evolution and the likelihoods used. More precisely, we obtain $\Delta \Omega_M \sim 0.007$ and $\Delta \Omega_M \sim 0.009$ when including \textit{Pantheon} and \textit{Pantheon +} samples, respectively, and $\Delta H_0 \sim 0.14$ with both SNe Ia samples (see Table \ref{tab:bestfit}). On the contrary, \citet{snelikelihood2022} have achieved a significant reduction of the uncertainties of a factor $\sim 40\%$ by applying the proper likelihood to SNe Ia instead of the traditional Gaussian one. Indeed, \citet{snelikelihood2022} have used only SNe Ia, while we here consider the combination of SNe Ia, GRBs, QSOs, and BAO, increasing the sample of SNe Ia with 2497 additional sources.
    The combination of different probes and the increase in the number of data, compared to the use of SNe Ia alone, already reduces the uncertainties up to a value lower than the one obtained with the new likelihoods in \citet{snelikelihood2022}. Specifically, using the new proper likelihoods they reach $\Delta \Omega_M = 0.012$ and $\Delta H_0 = 0.20$ with the \textit{Pantheon} sample, and $\Delta \Omega_M = 0.011$ and $\Delta H_0 = 0.16$ with the \textit{Pantheon +} sample. In our analysis, we achieve $\Delta \Omega_M \sim 0.08$ and $\Delta H_0 = 0.15$ just by combining all the probes with the standard $\mathcal{L}_{Gaussian}$. Hence, the increment in the number of data (i.e. 3545 with \textit{Pantheon} and 4198 with \textit{Pantheon +}) and the combination of different probes are the principal causes of the reduction on the uncertainties of cosmological parameters, making the choice of the proper likelihood of secondary impact. 
    However, as proved by the results of \citet{snelikelihood2022}, the search for the proper likelihood still represents a crucial point to validate the cosmological analysis on a statistical basis and for precision cosmology, in particular when dealing with new probes, small data sets, different combinations of probes, and new samples. 
    We here also stress that, between the two primary factors,
    the use of independent probes has the strongest impact. In this regard, the addition of BAO has the most relevant effect. Indeed, \citet{2022MNRAS.tmp.2639D} have already shown that adding BAO to \textit{Pantheon} SNe Ia lowers the uncertainties up to $\Delta \Omega_M = 0.07$ and  $\Delta H_0 = 0.14$ (see their Table 5). Nevertheless, increasing the number of cosmological data sets could still play a significant role in future analyses, depending also on the cosmological probes investigated. 
    We here also would like to stress that the difference between the uncertainties on $H_0$ that we obtain and the ones reported in \citet{2019ApJ...876...85R} for \textit{Pantheon} and \citet{2022ApJ...934L...7R} for \textit{Pantheon +} are due to the different approaches used to determine $H_0$. Indeed, \citet{2019ApJ...876...85R} and \citet{2022ApJ...934L...7R} use Cepheids in SNe Ia hosts, Cepheids as anchors or non–SNe Ia hosts, SNe Ia in Cepheids hosts, external constraints, and SNe Ia in the Hubble flow to determine, among all the free parameters of the fit, $5 \, \mathrm{log_{10}}H_0$, and hence $H_0$ and its uncertainty (by also adding a contribution to the systematic uncertainty from the analysis variants). The value of $H_0$ presented in \citet{2022ApJ...934L...7R} is then used (with its corresponding $M$) to compute the distance moduli and the corresponding uncertainties both supplied by the \textit{Pantheon +} release. On the other hand, the \textit{Pantheon} release provides the distance moduli with the arbitrary assumption of $H_0= 70 \, \mathrm{km} \, \mathrm{s}^{-1} \, \mathrm{Mpc}^{-1}$. In this work instead we consider the whole SNe Ia sample (1048 sources for \textit{Pantheon} and 1701 sources for \textit{Pantheon +}) and not only 42 SNe Ia by fitting directly the distance moduli and their uncertainties, which are provided by the \textit{Pantheon} and \textit{Pantheon +} releases, as described in Section \ref{quantities}. Thus, since in the values of the distance moduli the values of $M$ are fixed in correspondence of either $H_0= 70 \, \mathrm{km} \, \mathrm{s}^{-1} \, \mathrm{Mpc}^{-1}$ and $H_0= 73.04 \, \mathrm{km} \, \mathrm{s}^{-1} \, \mathrm{Mpc}^{-1}$ for the \textit{Pantheon} and \textit{Pantheon +} sample respectively, this leads to uncertainties on $H_0$ that are significantly reduced compared to the ones reported in \citet{2019ApJ...876...85R} and \citet{2022ApJ...934L...7R}.

    \item Independently on the approach of the likelihood and correction for evolution, we can identify a specific trend when comparing the two cases with the inclusion of \textit{Pantheon} and \textit{Pantheon +} samples provided in Table \ref{tab:bestfit}. This trend has already been pointed out in \citet{snelikelihood2022} through a deep investigation of the assumptions on which the two releases rely. Indeed, as in their work, we find that the inclusion of SNe Ia from \textit{Pantheon} results in $\Omega_M \sim 0.3$ and $H_0 = 70 \, \mathrm{km} \, \mathrm{s}^{-1} \, \mathrm{Mpc}^{-1}$, while considering the \textit{Pantheon +} sample we obtain $\Omega_ M \sim 0.36$ and $H_0 = 73 \, \mathrm{km} \, \mathrm{s}^{-1} \, \mathrm{Mpc}^{-1}$. More precisely, the values we obtain for $\Omega_M$ and $H_0$ show a discrepancy always greater than 3 $\sigma$ between the data sets with \textit{Pantheon} and with \textit{Pantheon +} SNe Ia, with the largest inconsistency being 10 $\sigma$ in the $H_0$ value (see Table \ref{tab:bestfit}). As stressed in \citet{snelikelihood2022}, this difference can be ascribed to the a-priori assumption on $M$ in Eq. \eqref{muobs} for the two data sets. As already detailed in Section \ref{quantities}, \citet{scolnic2018} assumes $M=-19.35$ corresponding to $H_0 = 70 \, \mathrm{km} \, \mathrm{s}^{-1} \, \mathrm{Mpc}^{-1}$, while \citet{pantheon+} uses $M=-19.253$ corresponding to $H_0 = 73.04 \, \mathrm{km} \, \mathrm{s}^{-1} \, \mathrm{Mpc}^{-1}$, as obtained in \citet{2022ApJ...934L...7R}. Since $\Omega_M$ and $H_0$ are related through Eq. \eqref{dmlcdm_corr}, an assumption on $H_0$ directly impacts the constraints on $\Omega_M$. 
    
    \item Our best-fit results and the ones presented in \citet{snelikelihood2022} are completely compatible (see Table \ref{tab:bestfit}) even if we are adding GRBs, QSOs, and BAO. Indeed, GRBs and QSOs alone cannot constrain cosmological parameters, while BAO provide best-fit values of the cosmological parameters compatible in 1 $\sigma$ with the ones of SNe Ia, but with larger uncertainties \citep{2017ApJ...849...84W,2019A&A...629A..85D,2022MNRAS.515.1795B}, due to the few number of data. Thus, in our combination of four cosmological probes, SNe Ia play the leading role in the determination of $\Omega_M$ and $H_0$.

    \item 
    Similarly to \citet{snelikelihood2022}, we obtain $\zeta_{P+} \sim 3$ when it is computed for the values obtained with \textit{Pantheon}, and $\zeta_P \sim 15$ when calculated for the data set with \textit{Pantheon +} SNe Ia, mainly due to the error on the reference value (i.e. 0.13) which is much lower than the errors of the other two $H_0$ reference values (i.e. 0.5 for $\zeta_{CMB}$ and 1.04 for $\zeta_{P+}$). Concerning $\zeta_{CMB}$, it is $\sim 5$ and $\sim 10$ when considering \textit{Pantheon} and \textit{Pantheon +} samples, respectively, since the $H_0$ obtained in \citet{scolnic2018} is less discrepant from the one in \citet{planck2018} compared to $H_0$ in \citet{2022ApJ...934L...7R}.
    In addition, we also stress that all our $H_0$ values are compatible with the one obtained in \citet{2021ApJ...919...16F} from the TRGB, which is $H_0=69.80 \pm 1.60 \, \mathrm{km} \, \mathrm{s}^{-1} \, \mathrm{Mpc}^{-1}$ when including systematic uncertainties. Specifically, the consistency is within 1 $\sigma$ when considering the \textit{Pantheon} SNe Ia, and within 2 $\sigma$ when using the \textit{Pantheon +} sample.

    \item Comparing our results on $\Omega_M$ with $\Omega_M=0.315 \pm 0.007$ from \citet{planck2018}, $\Omega_M=0.298 \pm 0.022$ from \citet{scolnic2018}, and $\Omega_M=0.338 \pm 0.018$ from \citet{2022ApJ...938..110B}, we find that our $\Omega_M$ values are always compatible within 3 $\sigma$ with each of these values, with no dependence on the SNe Ia sample, likelihood, or correction for evolution considered. More specifically, if we use the \textit{Pantheon} sample, the compatibility with $\Omega_M=0.315 \pm 0.007$ is within 1 $\sigma$ with $\mathcal{L}_{Gaussian}$ and within 2 $\sigma$ with $\mathcal{L}_{new}$, while, using the \textit{Pantheon +} sample, the difference increases to 2--3 $\sigma$ for both likelihoods. Compared with $\Omega_M=0.298 \pm 0.022$, the consistency is within 1 $\sigma$ for the case with \textit{Pantheon} SNe Ia, and within 2 $\sigma$ with \textit{Pantheon +}. In relation to $\Omega_M=0.338 \pm 0.018$, both data sets show a 1--2 $\sigma$ compatibility.

    \item 
    We explore the QSO role in the overall data set by comparing our results with the ones in \citet{2022MNRAS.tmp.2639D}, in which the combination of the same samples of GRBs and BAO as in this work and SNe Ia from \textit{Pantheon} is used to constrain $\Omega_M$ and $H_0$ both without correction for evolution and with a fixed correction. We find compatibility in 1 $\sigma$ for the values of $\Omega_M$ and $H_0$.
    In addition, also the uncertainties on these parameters do not show any difference. This implies that QSOs are still too weak, mainly due to the intrinsic dispersion of the RL relation, to significantly contribute to constraining cosmological parameters, compared to the dominant power of SNe Ia and BAO. As a matter of fact, the same conclusion can be drawn for GRBs. Indeed, as shown in \citet{2022MNRAS.tmp.2639D}, the inclusion of GRBs in the BAO + \textit{Pantheon} SNe Ia data set does not impact the cosmological results, neither in the best-fit values nor in the associated uncertainties (see their Table 5). On the other hand, \citet{2022MNRAS.tmp.2639D} proved that adding BAO to SNe Ia
    significantly reduces the uncertainties on these parameters, as already stressed above. Thus, in the total data set used in this work, BAO and SNe Ia play the primary role in constraining the cosmological parameters. The non-predominant role of QSOs and GRBs in our analysis is also the reason why different treatments for the evolution in redshift of their physical quantities do not affect the results, leading to best-fit values of $\Omega_M$ and $H_0$ compatible within 1 $\sigma$, independently on the likelihood and SNe Ia sample considered. 
    Regarding the different approaches to treat the reshift evolution of QSOs and GRBs, we here also stress that we obtain completely consistent results both when using the functions $k(\Omega_M)$ and $k$ as free parameters (see Figures \ref{fig: Om+H0 var Ev}, \ref{fig: kfree Gaussian}, and \ref{fig: kfree newlikelihoods}). Indeed, not only $\Omega_M$ and $H_0$ are compatible within less than 0.1 $\sigma$, but also the best-fits of $k$ parameters are within less than 1 $\sigma$ compared to the ones expected from $k(\Omega_M$). This proves also the reliability of our method of varying evolution with the use of the functions $k(\Omega_M$).
    Besides their secondary role in this analysis, GRBs and QSOs are still very promising cosmological tools. Indeed, as the methodology to standardize them as cosmological probes has been developed only very recently, they cannot be as powerful as the traditionally used probes. Nonetheless, they manifest an incredible potential to significantly contribute to the cosmological analysis. First, as high-$z$ probes,
    they allow us to investigate a previously unexplored region in the Universe evolution, which is the one crucially needed to discern between predictions of different cosmological models, indistinguishable in low-$z$ range of SNe Ia. 
    Additionally, their power in constraining cosmological parameters shows significant margins of improvement under different points of view. We here list a few: an increase in the number of sources, new and higher-quality observations, a refining in the selection of the samples, that would lead to a reduced intrinsic dispersion of the Dainotti and RL relations (and thus to tighter constraints on cosmological parameters). 
     Indeed, as already anticipated in \citet{Dainotti11b} and \citet{dainotti11a}, the choice of a morphological well-behaved sample can lead to these probes as distance indicators and to reduced cosmological parameters uncertainties.
     Lastly, and very importantly new studies that can enhance further the understanding on their physical mechanisms and backgrounds will better validate their application in cosmology.

\end{itemize}

\section{Summary \& Conclusions}
\label{conclusions}

In this work, we have first performed a statistical investigation on BAO and the most updated cosmological samples of GRBs and QSOs to test if the commonly adopted approach to constrain cosmological parameters by using a Gaussian likelihood is actually legitimated. 
For GRBs and QSOs, we have also taken into account the evolution in redshift of the variables as done in \citet{2022MNRAS.tmp.2639D} and \citet{biasfreeQSO2022}. Following the method applied to SNe Ia in \citet{snelikelihood2022}, we have employed several independent normality tests and fitted the $\Delta$ histograms to find the best-fit distribution. Indeed, when fitting cosmological models, the use of the best-fit likelihood function for each probe considered is crucial not only to build a cosmological likelihood that is statistically well-founded and justified, but also to obtain reliable and intrinsic results and the smallest uncertainties on cosmological parameters \citep{snelikelihood2022}. As a second step in our analysis, we have fitted the flat $\Lambda$CDM model with the combination of SNe Ia, GRBs, QSOs, and BAO considering both \textit{Pantheon} and \textit{Pantheon +} SNe Ia samples, applying different approaches to treat the redshift evolution for GRBs and QSOs, and leaving both $\Omega_M$ and $H_0$ as free parameters. This fit is performed with two different methods: using Gaussian likelihoods for all the probes ($\mathcal{L}_{Gaussian}$), as in the usual practise in the literature, and considering for each probe the new likelihoods uncovered in this work ($\mathcal{L}_{new}$). 

Surprisingly, the statistical investigation on the Gaussianity assumption has revealed that only GRBs obey a Gaussian distribution, while QSOs and BAO show a logistic and a Student's t distribution, respectively. 
Despite this unexpected result, this analysis is significantly relevant, because the Gaussianity requirement is not the only requirement to be considered in enlarging the current statistical samples and probes.
Thus, the practise traditionally trusted in cosmological analyses and also future studies will have to be inspected carefully for new samples and probes. 

Concerning the fit of the flat $\Lambda$CDM model, we have shown that our cosmological results are not affected either by the approach used for the redshift evolution of GRBs and QSOs or by the choice of the likelihood. Indeed, in all the cosmological cases studied, the main role is played by SNe Ia and BAO. Indeed, SNe Ia are the leading probes that dominate above GRBs, QSOs, and BAO in driving the determination of $\Omega_M$ and $H_0$, as proved also by the analysis on the rescaling of $M$. On the other side, the inclusion of BAO represents the dominant factor in reducing the uncertainties on the cosmological parameters.
Specifically, we have obtained very tight constraints, up to $\Delta \Omega_M = 0.007$ and $\Delta H_0 = 0.13$. Our values of $\Omega_M$ are always compatible
with the values from \citet{scolnic2018}, \citet{planck2018}, and \citet{2022ApJ...938..110B}. Regarding $H_0$, our results are consistent
with the value obtained from the TRGB, while they show a significant discrepancy from the value derived from the CMB.

In conclusion, this work shows 
the importance of inspecting the Gaussianity assumption for all cosmological probes and future new samples in order to apply the most appropriate likelihood in constraining cosmological parameters.
We have also clearly stressed that the cosmological results on $\Omega_M$ and $H_0$ obtained with SNe Ia are induced by the a-priori calibration imposed on them. Bearing this in mind, we have found compatibility between our $H_0$ values and the one from the TRGB, and between our $\Omega_M$ values and the ones from \citet{planck2018}, \citet{scolnic2018}, and \citet{2022ApJ...938..110B}. 
Finally, we have also highlighted the relevance of the inclusion of GRBs and QSOs in cosmological studies and their bright potential in this field.
Indeed, our study is motivated by the current need of cosmological probes at intermediate redshift between the one of SNe Ia and CMB, and, to this end, QSOs and GRBs represent the most promising sources to date. Thus, we have here proved their applicability in cosmological analyses when they are used jointly with more powerful and well-established probes, such as SNe Ia and BAO. We are not at the stage in which we expect QSOs or GRBs to play the leading role in constraining cosmological parameter. Nevertheless, we have shown how they can be implemented in the cosmological analysis to extend the redshift range up 7.5 without introducing additional uncertainties on the cosmological parameters. This is the first essential step to improve their use in cosmology. In this regard, \citet{2022MNRAS.514.1828D} have already estimated the number of GRBs (and the time needed to observe them) to reach the same precision of \textit{Pantheon} SNe Ia with GRBs. Similar efforts in determining a sub-sample of QSOs that even now provides constraints on $\Omega_M$ with a precision comparable with the one obtained from \textit{Pantheon + } SNe Ia is underway.

\section*{Acknowledgements}

This study uses data supplied by SNe Ia GitHub repositories \url{https://github.com/dscolnic/Pantheon} and \url{https://github.com/PantheonPlusSH0ES}.
GB acknowledges the Istituto Nazionale di Fisica Nucleare (INFN), sezione di Napoli, for supporting her visit at NAOJ. GB is grateful to be hosted by Division of Science. MGD acknowledges the Division of Science and NAOJ. SN acknowledges JSPS KAKENHI (A: 19H00693), Interdisciplinary Theoretical and Mathematical Sciences Program (iTHEMS), and the Pioneering Program of RIKEN for Evolution of Matter in the Universe (r-EMU). SC acknowledges Istituto Nazionale di Fisica Nucleare (INFN), sezione di Napoli, {\it iniziativa specifica} QGSKY.
This paper is partially based upon work from COST Action CA21136 {\it Addressing
observational tensions in cosmology with systematics and fundamental
physics} (CosmoVerse) supported by COST (European Cooperation in Science and
Technology).
We are particularly grateful to H. Nomura for the discussion about the Gaussianity results on GRBs. We also acknowledge B. De Simone for his help in launching some of the notebooks to produce the results with the evolutionary parameters free to vary.

\section*{Data Availability}

The data underlying this article will be shared upon a reasonable request to the corresponding author.



\bibliographystyle{mnras}
\bibliography{bibliografia} 




\appendix


\bsp	
\label{lastpage}
\end{document}